\documentclass{aa}  
\usepackage[dvipsnames]{xcolor}
\usepackage{graphicx}
\usepackage{txfonts}
\usepackage{xcolor}
\usepackage{mathrsfs}
\usepackage[colorlinks=true, citecolor=blue]{hyperref}
\usepackage{physics}
\usepackage{nicefrac}
\usepackage[normalem]{ulem}
\usepackage{soul}
\usepackage[colorlinks=true, citecolor=blue]{hyperref}

\usepackage[version=4,arrows=pgf-filled,
textfontname=sffamily,
mathfontname=mathsf]{mhchem}

\newcommand{\vr}{\vec{r}}

\newcommand{\TQ}{Q_{\text{bi-fluid}}}

\newcommand{\Tcd}{\Tilde{c}_d}
\newcommand{\Tcg}{\Tilde{c}_g}

\newcommand{\Txi}{\Tilde{\xi}}
\newcommand{\Tz}{\Tilde{z}}

\newcommand{\betaf}{\text{B}}
\newcommand{\Tf}[2]{\Theta_{#1}\left(#2\right)}

\newcommand{\az}{\alpha_z}
\newcommand{\scz}{\text{Sc}_z}
\newcommand{\sto}{\text{St}}

\begin{document}

\title{Gravito-turbulent bi-fluid protoplanetary discs:}
\subtitle{1. An analytical perspective to stratification}

\author{S. Rendon Restrepo \inst{1}
\thanks{\email{srendon@aip.de}}
\and
U. Ziegler \inst{1}
\and
M. Villenave \inst{2}
\and
O. Gressel \inst{1}
}
\institute{Leibniz-Institut für Astrophysik Potsdam (AIP), Potsdam, Germany
\and Universitá degli Studi di Milano, Dipartimento di Fisica, via Celoria 16, 20133 Milano, Italy}


 
\abstract
{In Class 0/I as well as the outskirts of Class II circumstellar discs, the self-gravity of gas is expected to be significant, which certainly impacts the disc vertical hydrostatic equilibrium.
The contribution of dust, whose measured mass is still uncertain, could also be a factor in this equilibrium.
}
{We aim to formulate and solve approximately the equations governing the hydrostatic equilibrium of a self-gravitating disc composed of gas and dust.
Particularly, we aim to provide a fully consistent treatment of turbulence and gravity, affecting almost symmetrically gas and dust.
From an observational perspective, we study the possibility of indirectly measuring disc masses through gas layering and dust settling measurements. 
}
{We used analytical methods to approximate the solution of the 1D Liouville equation with additional non-linearities governing the stratification of a self-gravitating protoplanetary disc. 
The analytical findings were verified through numerical treatment and their consistency validated with physical interpretation.
}
{For a constant vertical stopping time profile, we discovered a nearly exact layering solution valid across all self-gravity regimes for gas and dust. 
From first principles, we defined the Toomre parameter of a bi-fluid system as the harmonic average of its constituents' Toomre parameters. 
Based on these findings, we propose a method to estimate disc mass through gas or dust settling observations. 
We introduce a generic definition of the dust-to-gas scale height, applicable to complex profiles. 
Additionally, we identified new exact solutions useful for benchmarking self-gravity solvers in numerical codes.
}
{The hydrostatic equilibrium of a gas/dust mixture is governed by their Toomre parameters and their effective relative temperature. 
The equilibrium we found could be used for possibly measuring disc masses, enabling a more thorough understanding of disc settling and gravitational collapse, and will also improve the computation of self-gravity in thin disc simulations.
}

\keywords{self-gravity --
          bi-fluid --
          hydrodynamics --
          Protoplanetary discs}

\maketitle
%

\section{Introduction}

Protoplanetary accretion discs (PPDs) get their name from the process of material infalling into their host star.
This accretion mechanism is associated with the outward transport of angular momentum, a process thought to originate from magnetised disc winds or a possible source of turbulence.
The latter process can be driven by one of the main families of instabilities: magnetohydrodynamical instabilities \citep{1991_balbus}, thermo-hydrodynamical instabilities, among them the vertical shear instability, convective overstability and zombie vortex instability \citep{2013_nelson, 2014_klahr,2005_barranco, 2013_marcus}.
We can also include resonant drag instabilities \citep{2018_squire_hopkins}, which arise from aerodynamic interactions between gas and dust.
A particular case of these instabilities is the notable streaming instability \citep{2005_youdin, johansen_youdin_2007}.
Finally, young discs or the outer regions of Class I/II discs are subject to the Gravitational Instability (GI) \citep{2001_gammie,2016_Kratter_Lodato}.

The measure of accretion rates, ranging in $ 10^{-10}-10^{-7} \, M_{\odot}$/yr \citep{2014_venuti,2016_hartmann,2016_manara}, is one of the possible ways to estimate the intensity of turbulence \citep{1998_hartmann}.
Turbulence strength in PPDs can also be determined through Doppler broadening of spectral data, molecular emission of CO lines \citep{2004_carr,2020_flaherty}, and by observing the settling of dust particles in the disc midplane.
Indeed, particle settling, promoted by the star's gravity, is balanced by ascending turbulent transport, which allows to relate the dust-to-gas scale height ratio to the vertical diffusion and Stokes number \citep{1995_dubrulle}.
From millimeter observations, most Class II discs exhibit significant level of dust settling in their outer discs, among them HL Tau, Oph163131 and HD163296~\citep{2016_pinte, 2020_villenave, 2022_villenave, Doi_Kataoka_2021, Pizzati_2023}, indicating very low levels of turbulence.
Therefore, it seems evident that the incorporation of additional relevant effects to the vertical hydrostatic equilibrium of the disc will allow finer gas and dust stratifications, possibly permitting to access more properties from the disc.
One of these additional ingredients could be the gravity of the disc on itself, colloquially known as self-gravity (SG).

When the disc SG is insignificant compared to the star's gravity gas and dust adopt a Gaussian vertical profile \citep[Eqs. 15 and 238]{2019_armitage}, a situation that is assumed for measuring the dust-to-gas scale height ratio in class I/II discs \citep[see Sect. 2.3.1]{2023_miotello_ppvii}.
Conversely, in Class 0/I discs, the disc SG dominates. 
This is typically seen in systems where mass infall is still ongoing, such as the Elias 2-27 system \citep{2016_perez,2021_veronesi}, or in the cold outer regions of PPDs, like in the RU Lup system \citep{2020_huang}.
For such a case, it was found that the gas vertical density stratification rather obeys a hyperbolic secant profile corresponding to the limit of small Toomre's parameter \citep{1942_spitzer, 1999_Bertin_Lodato}.
Same profile was also obtained in a dust rich environment by \citet{2020_klahr_schreiber, 2021_klahr_schreiber}, where they additionally defined a dust Toomre's parameter.
Aforementioned profiles however consider independent contributions of gas and dust SG to their own respective stratification. 
In other words, from a gravitational point of view, gas interacts solely with gas and dust interacts solely with dust, a scenario that might hold true in certain instances but is not generally applicable.
A more sophisticated model involves taking into account the star's gravity and the disc SG by using a biased Gaussian stratification, where all SG information is incorporated into a modified scale height, which is naturally Toomre's parameter dependent \citep{1999_Bertin_Lodato}.
This improvement allows for a smooth transition between Keplerian discs and massive discs.
The most comprehensive model to date, developed by \citet{2021b_baehr_zhu}, attempted to capture the effect of the star's gravity and the disentangled SG of gas and dust into the solid material.

It's worth noting that existing analytic models of discs layering that account for dust stirring in a turbulent gas environment often disregard the effect of turbulence on the gas itself.
In reality, turbulence should act as a pressure term for the gas \citep{1987_bonazzola}.
Therefore, to ensure consistency with the assumption of dust embedded in a turbulent gaseous environment, it is important to also assume that the gas itself is subject to turbulence.
This could permit to improve the models of \citet{1995_dubrulle}, \citet{2002A_takeuchi}, \citet{2009b_fromang_nelson} and \citet{2021_klahr_schreiber} where gas is considered as a passive tracer that kept its Gaussian stratification and its scale height, which is not necessarily true in presence of turbulence, and even less so when there is SG. 
This is confirmed by \citet{2021b_baehr_zhu} study, where they showed, thanks to 3D shearing box simulations, that the consideration of gas and dust SG on particles has significant consequences on the gas and dust settling.
Thus, more sophisticated models are needed for capturing the turbulence and SG affecting, both fluids, gas and dust.

In a different context, which was the initial focus of this work, accurately estimating gas and dust scale heights was found to significantly impact the calculation of self-gravity forces in thin disc (2D) simulations. 
These simulations typically use a Plummer potential with a smoothing length for gravity prescription \citep{muller_kley_2012}. 
\citet{rendon_restrepo_2023} demonstrated that while this method correctly models long-range SG interactions, it underestimates mid/short-range interactions by up to 100\%, aligning with the removal of Newtonian behaviour due to softening \citep{1989_adams, hockney2021computer}. 
Conversely, without smoothing, SG is highly overestimated. 
To address this, a space-varying smoothing length was introduced to better match the 2D self-gravity force with its 3D counterpart.
Further, \citet{rendon_restrepo_2023} extended their correction to bi-fluids with an embedded dust phase, finding that two additional smoothing lengths are needed to account for dust-dust and dust-gas gravitational interactions. 
Their aforementioned length depends heavily on the scale heights of gas and dust, making the vertical layering of a PPD crucial for accurate SG computation in 2D simulations. 
This is particularly important for massive discs, where consistency between vertical layering and in-plane SG computation is required.

In this first article of a series of two, we study the stratification of gas and dust in presence of turbulence and their disentangled SG contributions by analytical means. 
In the second accompanying paper, we study thanks to 3D hydrodynamical shearing box simulations how these results could potentially affect planetesimal formation and gas accretion.
We begin by detailing our assumptions and establishing the equations of hydrostatic equilibrium for gas and dust in Sect. \ref{Eq: Formulating the main equations}.
In Sects. \ref{sec: stopping time with vertical profile and negligible dust mass}-\ref{sec: constant stopping time and contribution of dust} we propose, depending on the assumption made on the vertical profile of the stopping time, different exact or general approximated solutions valid for any gravity contribution strength, i.e from  the star, gas or dust.
In Sect. \ref{sec: Connecting theory to observations: indirect mass estimation} we show how our results could permit to access indirectly the mass content of PPD.
Finally, in Sect. \ref{sec: discussion} we propose a discussion on the limitations of our approach, the possible consequences for planetesimal formation and the consequences for SG computation in thin disc simulations.
Important equations and detailed derivations can be found in appendices.

\section{Formulating the main equations} \label{Eq: Formulating the main equations}

In this Sect. we derive the two main equations describing the vertical hydrostatic equilibrium of gas and steady-state transport of dust in presence of turbulence and SG.
But first we introduce our notation conventions.

\subsection{Guidelines and cautions}
\label{subsec: guidelines and cautions}

We account for turbulence as an additional pressure term in the gas and a diffusion term for dust.
Both give rise to effective sound speeds, which we denote by a tilde: ( \textbf{$\Tilde{}$} ), e.g. $\Tcg$ stands for the modified gas sound speed accounting for turbulence.
The vertical stratification of a self-gravitating disc is a challenging problem that requires a redefinition of several physical quantities.
For instance, intuitively we expect the vertical scale heights to be reduced, which could be interpreted as a greater differential rotation in the midplane.
Therefore the differential rotation in presence of SG is also modified, which, surprisingly, changes the definition of the Toomre parameters as well.
Therefore, in order to avoid confusion we will use the upper-script \textbf{sg} as soon as a quantity is redefined in presence of SG.
Furthermore, the lower-scripts \textbf{g} or \textbf{d} will be used to make a distinction between gas and dust quantities, respectively.
If there is no special superscript, it means that we are using the reference notation used in the literature.
All these cautions will be particularly important for Sects. \ref{sec: stopping time with vertical profile and negligible dust mass} and \ref{sec: constant stopping time and contribution of dust}.
Finally, the "mid" subscript indicates that quantities are taken in the equatorial plane, $z=0$. 
In Table \ref{tab:definitions} we collected the definition of all quantities used in this study.

\subsection{Hydrostatic equilibrium of gas}

The gas is assumed to be in a turbulent situation.
Accordingly, we employ the Reynolds averaging method \citep{1895_reynolds, 1974_krause} and decompose all velocity components of gas as the sum of a mean and a fluctuating part: $v_i = \overline{v_i} + v_i'$.
The bar designates a time average.
A proper treatment of turbulence would require to decompose also the pressure and density variables into a mean and fluctuating part, and to consider the compressibility of the flow.
Using Favre averages is a more appropriate approach for this situation \citep{1965_favre, 1992_favre}, but it complicates the definition of diffusion terms, making it challenging to compare our work with existing literature that uses Reynolds averaging.
Consequently, in this work we assume that pressure and density are equal to their statistical mean value.
Despite this simplifying assumption, our model still represents an improvement over other models that ignore the effects of turbulence on the gas itself, such as those proposed by \citet{2009b_fromang_nelson} and \citet{2021b_baehr_zhu}.
By taking turbulence into account solely through velocity fluctuations, we keep the problem analytically tractable.
Assuming hydrostatic equilibrium, $\overline{v_z}=0$, and that the density, pressure, and turbulent $\overline{v_i' v_j'}$ terms are only $z$-dependent, we obtain the Reynolds equation for the gas in the vertical direction:
\begin{equation}
\frac{1}{\rho_g} \partial_z \left( p + \overline{\rho_g v_z' v_z'} \right) = \overline{F_z^g} 
\end{equation}
where $F_z^g$ represents all volume forces applied to the gas, which will be discussed in more detail in Sect. \ref{subsec: hydrostatic equilibrium equations of gas and dust}.
The gas is assumed to be in a vertical isothermal state, $p=c_g^2 \rho_g$, where $c_g$ is the isothermal sound speed.
The Reynolds stress term, $\overline{v_z' v_z'}$, can be rewritten in terms of a diffusion term in the $\alpha$-viscosity paradigm \citep{shakura_1973}: 
\begin{equation}
\az
= \frac{\overline{\rho_g v_z' v_z'}}{\overline{c_g^2 \rho_g}}
\end{equation}
For simplifying our approach, we assume that above turbulent diffusivity is constant with respect to $z$.
This assumption, which in general is not true \citep{2013_nelson,2020_flock, 2021b_baehr_zhu}, allows the problem to be accessible analytically.
It is important to note that in a majority of numerical and theoretical studies, the radial Reynolds stress, $\alpha_S={\overline{\rho_g v_r' v_\phi'}}/{ \overline{c_g^2 \rho_g} }$, responsible for transporting angular momentum in the radial direction, is frequently employed in place of the vertical stress. 
This stress is related to its vertical analogue through the equation $\az=\alpha_S/\scz$, where $\scz$ is the vertical Schmidt number, which quantifies the level of anisotropy in turbulence.
In the context of a PPD, the parameter $\az$ can take different values depending on the region of the disc, its age and most importantly on the type of instability generating the turbulence \citep{2023_PPVII_chap13}.
Introducing the effective gas sound speed, $\Tilde{c}_g=\sqrt{ 1 + \az } \, c_g$, the vertical hydrostatic equilibrium for gas can be rewritten as:
\begin{equation}\label{Eq: hydrostatic equation gas}
\Tilde{c}_g^2 \partial_z \ln \left( \rho_g \right)                
= \overline{F_z^g} 
\end{equation}

\subsection{Transport equation for dust}

The dust stirring stems from aerodynamic coupling of small grains, $\Omega_K \tau_f < 1$, with the turbulent gaseous environment, which allows to make use of the transport equation for dust \citep[Eq. 28]{1995_dubrulle, 2004_schrapler_henning}:
\begin{equation}
\partial_t \rho_d  + \partial_z \left[ \tau_f \rho_d F_z^d - \rho_g \kappa_t \partial_z \left( \frac{\rho_d}{\rho_g} \right) \right] 
=
0
\end{equation}
where $F_z^d$ are all volume forces applied to dust, except the aerodynamic force (see Sect. \ref{subsec: hydrostatic equilibrium equations of gas and dust}).
The third term on the left hand side represents the diffusion flux of solid material.
The stopping time and turbulent diffusivity are depicted by $\tau_f$ and $\kappa_t$, respectively.
Assuming dust to be in a steady state in the vertical direction, we get:
\begin{equation}\label{Eq: dust hydrostatic}
c_d^2 \partial_z \ln \left( \frac{\rho_d}{\rho_g} \right) = F_z^d
\end{equation}
where $c_d=\sqrt{\kappa_t/\tau_f}$ can be interpreted as a dust sound speed.
This stems from a direct analogy with the vertical hydrostatic equilibrium of gas (Eq. \ref{Eq: hydrostatic equation gas}).
However, this comparison, although direct, remains rather empirical. 
Therefore, we have decided to summarise, adapt, and generalise the approach proposed by \citet[Appendix B]{2021_klahr_schreiber} in the next paragraph, as it is more rigorous.

First, we redefine the velocity of the solid material as $\vec{v}_d^*=\vec{v}_d-\kappa_t \frac{\rho_g}{\rho_d} \vec{\nabla} \left( \frac{\rho_d}{\rho_g} \right)$ and propose a new momentum equation associated with the evolution of this velocity field.
The revised momentum equation resembles the momentum equation of dust but includes an unknown source term, $\mathcal{S}$, that drives diffusion and must be determined.
By employing the procedure of the previously mentioned authors, we obtained the expression of the source term as:
\begin{equation}
\mathcal{S} = - \frac{\kappa_t}{\tau_f} \partial_z \ln{\left(\frac{\rho_d}{\rho_g}\right)}
\end{equation}
This term acts as a vertical gradient of a pressure-like term for dust, allowing Eq. \ref{Eq: dust hydrostatic} to be interpreted as the hydrostatic equilibrium between the pressure of dust and the vertical forces promoting dust settling.
The derived sound speed, $c_d$, differs from the root mean square (rms) velocity of particles as discussed by \citet[Sect. 2]{2021_klahr_schreiber}.
We highlight that our expression for $\mathcal{S}$ differs from the one suggested by \citet[Eq. B11]{2021_klahr_schreiber}.
Indeed, they assumed that $\rho_g$ is a space constant, an assumption that we have relaxed to allow for greater generality.

\subsection{The vertical hydrostatic equilibrium equations of a gas and dust self-gravitating turbulent disc}\label{subsec: hydrostatic equilibrium equations of gas and dust}

In this work we consider the vertical equilibrium of a PPD made of gas and dust and orbiting around a central object.
For the sake of simplicity we neglect the effect of magnetic pressure, ${B^2}/{8 \pi}$.
We assume that both species are subject to the star's gravity, the disc own gravity, which is commonly referred to as self-gravity (SG), and their mutual aerodynamic interaction.
The body forces for gas and dust are given by:
\begin{equation}\label{Eq: equality forces vertical direction}
\overline{F}_z^g = F_z^d = -\Omega_K^2 z - \partial_z \Phi_{disc}
\end{equation}
where $\Omega_K$ stands for the Keplerian frequency and $\Phi_{disc}$ is the gravitational potential of the disc made of gas and dust.
In the limit of a flattened system the vertical component of SG prevails over other components and we can make use of the Paczynski's, or slab, approximation \citep[Sect. 2.3.3]{1942_spitzer, 1978A_paczynski,2008_binney_tremaine}:
\begin{equation}\label{Eq: slab approx}
\partial_{zz}^2 \Phi_{disc} = 4 \pi G \left(\rho_g + \rho_d \right)
\end{equation}
We note that within this assumption the variation of the potential at a given position is solely influenced by the mass distribution at that specific location, rather than the mass distribution throughout the entire disc.
Putting together Eqs. \ref{Eq: hydrostatic equation gas}, \ref{Eq: dust hydrostatic}, \ref{Eq: equality forces vertical direction} and \ref{Eq: slab approx}, the stratifications of gas and dust are coupled via:
\begin{equation}\label{Eq: SG of gas and dust}
\left\{
\begin{array}{cc}
\displaystyle \tilde{c}_g^2 \partial_z \ln{\left(\rho_g\right)}          &= \displaystyle - \Omega_K^2 \, z - {2\pi G} \left( \Sigma_g(\vr,z) + \Sigma_d(\vr,z) \right) \\ [8pt]
\displaystyle c_d^2 \partial_z \ln{\left(\rho_d / \rho_g\right)} &= \displaystyle - \Omega_K^2 \, z - {2 \pi G} \left( \Sigma_g(\vr,z) + \Sigma_d(\vr,z) \right)
\end{array}
\right.
\end{equation}
where $\Sigma_i(\vr,z)=\int\limits_{-z}^{z} \rho_i(\vr,z') \, dz'$ is the vertically integrated density of gas and dust and $\vr$ is the position vector in polar coordinates.
Above equation expresses that gas, resp. dust, stratification is set by the balance between the star's and PPD SG and pressure, resp. turbulent vertical diffusion.
The Eq. \ref{Eq: SG of gas and dust} requires additional constraints, which in present case are obtained self-consistently thanks to the definition of the column densities: 
\begin{equation}\label{Eq: column density}
\Sigma_i(\vr) = \Sigma_i(\vr, z=\infty) = \displaystyle \int_{-\infty}^{\infty} \rho_i(\vr, z') \, dz' 
\end{equation}
This closure condition is particularly interesting since the disc mass, and thus indirectly column densities, are the quantities that can be probed thanks to optically thin tracers, 
like \ce{C ^{18}O} or \ce{N2H+} \citep{2021_zhang, 2022_trapman}.
Finally, the additional constraints lie in the assumptions made on the stopping time.

In the following sections, we'll initially address the broader scenario of a stopping time with a vertical profile and then proceed to the simpler assumption of a constant stopping time.

%

\begin{table}
\caption{Definitions}             
\label{tab:definitions}      
\centering          
\begin{tabular}{l l}     
\hline\hline       
Name                               & Symbol   + Definition  \\ 
\hline
\hline 
\multicolumn{2}{c}{\textbf{General}} \\
\hline 
Keplerian frequency                & $\Omega_K$            \\
Stopping time                      & $\tau_f$              \\
Stokes number                      & $\sto=\tau_f \Omega_K$  \\
Vertical turbulent diffusivity     & $\kappa_t$            \\
Dimensionless in plane $\alpha$-viscosity        & $\alpha_S$              \\
Dimensionless vertical $\alpha$-viscosity & $\displaystyle \az={\kappa_t \Omega_K}/{c_g^2}$ \\
Vertical Schmidt number            & $\scz=\alpha_S/\az$ \\  [4pt]
\hline 
\multicolumn{2}{c}{\textbf{Gas}} \\
\hline 
Gas sound speed                    & $c_g$ \\
Effective gas sound speed          & $\Tilde{c}_g = \sqrt{1+\alpha_z} c_g$ \\
Surface density of gas             & $\Sigma_g$ \\
Gas Toomre's parameter             & $Q_g=\frac{\Tilde{c}_g \Omega_K}{\pi G \Sigma_g}$ \\
Gas turbulent pressure scale height& $H_g = \Tilde{c}_g / \Omega_K$   \\ [4pt]
Gas gravito-turbulent pressure     & $H_g^{sg}$ (Eq. \ref{Eq: parameters stratification gas and dust} ) \\ 
scale height                       & \\  [4pt]
\hline 
\multicolumn{2}{c}{\textbf{Dust}} \\
\hline & \\[-1.5ex]
Dust sound speed                   & $c_d=\sqrt{{\kappa_t} / {\tau_f}}$ \\
Midplane dust sound speed          & $c_{d,{\rm mid}} = \sqrt{\frac{\az}{\sto}} c_g$ \\
Effective dust sound speed         & $\Tilde{c}_d = \frac{\xi}{\sqrt{1+\xi^2}} c_{d, {\rm mid}}$ \\
Surface density of dust            & $\Sigma_d$  \\
Effective dust Toomre's parameter  & $Q_d=\frac{\Tilde{c}_{d} \Omega_K}{\pi G \Sigma_d}$ \\
Dust diffusive scale height        & $H_d=\Tcd/\Omega_K$   \\ [4pt]
Dust gravito-diffusive scale height& $H_d^{sg}$ (Eq. \ref{Eq: parameters stratification gas and dust} )  \\ [4pt]
\hline 
\multicolumn{2}{c}{\textbf{Gas and dust combined}} \\
\hline & \\[-1.5ex]
Gas-to-dust temperature   & $\xi = \frac{\Tilde{c}_g}{c_{d,{\rm mid}}} = \sqrt{\frac{(1+\az)}{\az} \sto} $ \\
Effective gas-to-dust temperature  & $\Tilde{\xi} = \frac{\Tilde{c}_g}{\Tilde{c}_{d,{\rm mid}}}=\sqrt{1+\xi^2} $ \\ 
3D Toomre's parameter              & $Q_{i}^{3D} = \sqrt{ \frac{\pi}{2} } \frac{\Omega_{K}^2}{\pi^2 G \rho_{i,{\rm mid}}}$ \\
General Toomre's parameter         & $\TQ^{3D}=\left( \frac{1}{Q_g^{3D}} + \frac{1}{Q_d^{3D}} \right)^{-1}$ \\
for a turbulent bi-fluid           & \\
\hline                  
\end{tabular}
\end{table}

\section{Stopping time with vertical profile and negligible dust mass}
\label{sec: stopping time with vertical profile and negligible dust mass}


We will assume that the turbulent diffusivity coefficient $\kappa_t$ is constant but the stopping time has a density dependence:
\begin{equation}\label{Eq: def stopping time vertical profile}
\tau_f = \tau_{f, {\rm mid}} \frac{\rho_{g,{\rm mid}}}{\rho_g}
\end{equation}
For instance, this realistic consideration means that small solid material, ${\sto= \tau_f \Omega_K \ll 1}$, is less coupled to gas in the upper layers of the disc.
Further, we will assume in this Sect. that dust mass is negligible compared to gas mass.
Using Eq. \ref{Eq: def stopping time vertical profile}, it is convenient to reformulate the set of Eqs. \ref{Eq: SG of gas and dust} as a single equation for the unknown $\rho_g$ and a transformation which permits to retrieve $\rho_d$ from $\rho_g$:
\begin{equation}\label{Eq: reduced eq. SG}
\left\{
\begin{array}{lll}
\displaystyle \Tilde{c}_g^2 \partial_{zz}^2 \ln{\left(\rho_g\right)}          &= &\displaystyle - \Omega_K^2  - {4\pi G} \rho_g  \\ [8pt] 
\rho_d & = & \displaystyle A(\vr) \, \rho_g \, \exp\left( -\xi^2 \, \frac{\rho_{g,{\rm mid}}}{\rho_g} \right)
\end{array}
\right.
\end{equation}
with $\xi={\Tilde{c}_g}/{c_{d,{\rm mid}}} = \sqrt{\frac{(1+\az)}{\az} \sto}$.
The integration factor $A(\vr)$ is to be fixed by the constraint in Eq. \ref{Eq: column density}.

\subsection{An exact solution for massive gas discs}\label{subsec: An exact solution for massive gas discs}

\begin{figure}
\centering
\includegraphics[width=\hsize]{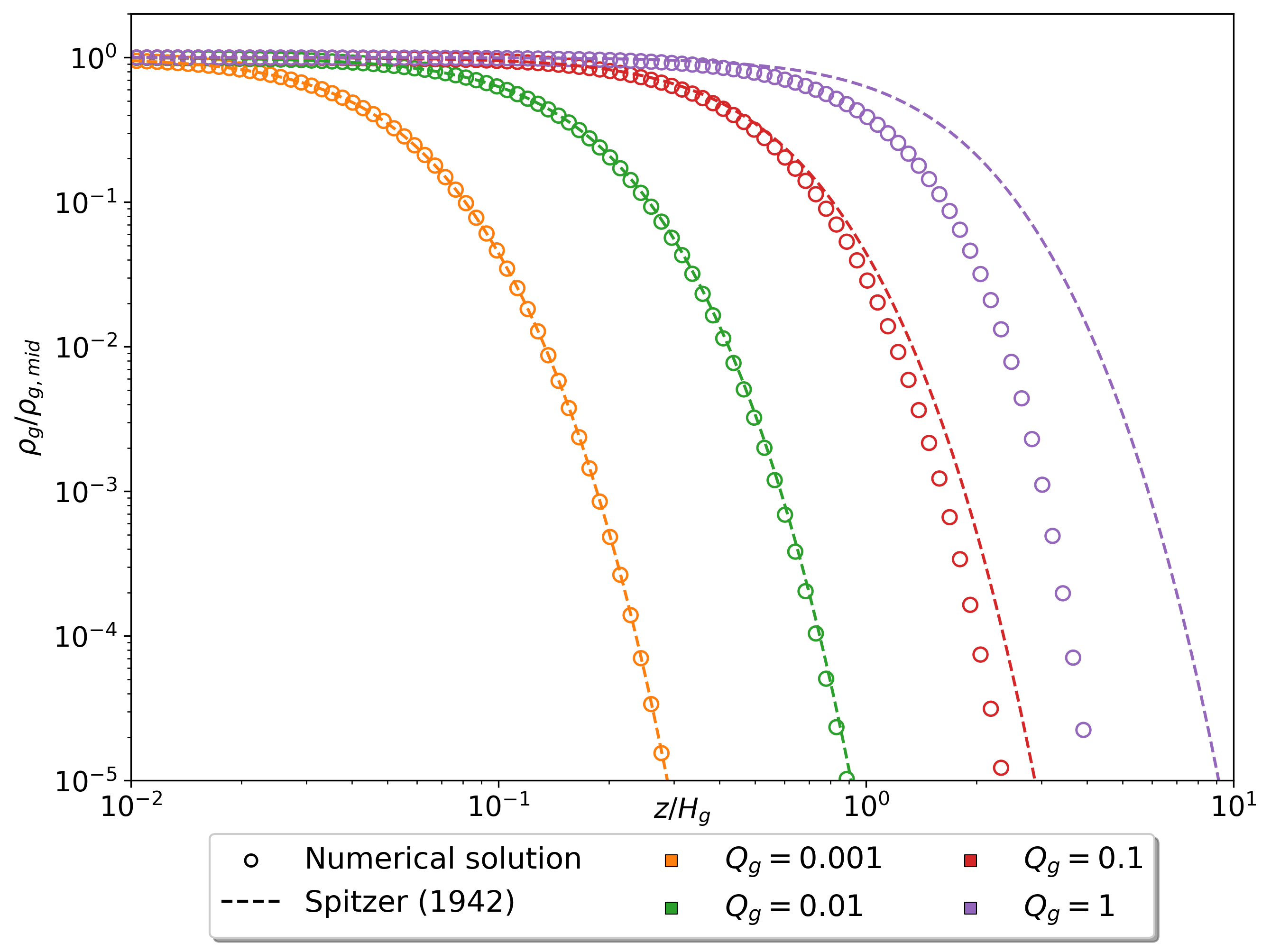}

\caption{Normalised gas density profile for various Toomre parameters ($Q_g$) comparing the numerical solution of Eq. \ref{Eq: reduced eq. SG} with the Spitzer solution.
}
\label{fig: numeric vs spitzer}
\end{figure}

In the case of a massive gas disc, i.e when $\rho_d \ll \rho_g$ and the gas Toomre parameter $Q_g=\frac{\Tcg \Omega_K}{\pi G \Sigma_g} \ll 1$, we found an exact solution to Eq. \ref{Eq: reduced eq. SG} given by:
\begin{equation}\label{Eq: No const. stopping time, exact high gas mass}
\left\{
\begin{array}{lll}
\rho_g(\vr,z)&=&\displaystyle \frac{\Sigma_g}{2 Q_g H_g}  \sech^{2}\left(\frac{z}{Q_g H_g}\right) \\ [10pt]
\rho_d(\vr,z)&=&\displaystyle \frac{\Sigma_d}{2 Q_g \exp\left(\xi^2\right) I_1(\xi) H_g} \sech^2\left( \frac{z}{Q_g H_g} \right)  \\ [10pt]
           & & \quad \exp\left[ - \xi^2 \left(\cosh^2\left(  \frac{z}{Q_g H_g} \right) - 1\right) \right]
\end{array}
\right.
\end{equation}
where $H_g=\Tcg/\Omega_K$ is the turbulent modified pressure scale height and
\begin{equation}
I_1(\xi) = \frac{\xi^2}{2} \exp\left( - \frac{\xi^2}{2} \right) \left[ K_1\left( \frac{\xi^2}{2} \right) - K_0\left( \frac{\xi^2}{2} \right) \right]
\end{equation}
with $K_\nu$ a modified Bessel function of the second kind and of order $\nu$.
The detailed derivation of the integration constant $A(r, \varphi)$ can be found in appendix \ref{app: non constant stopping time, only gas disc contribution}.
As expected, we retrieved the Spitzer solution for the gas layering \citep{1942_spitzer} and a more sophisticated profile for dust consisting of an envelope function (set by the gas vertical profile) and a function with rapid decline.
To our knowledge the solution for the dust profile is new.
It is worth mentioning that Eq. \ref{Eq: No const. stopping time, exact high gas mass} is the counterpart of \citet[Eq. 31]{2002A_takeuchi} and \citet[Eq. 19]{2009b_fromang_nelson} that is valid for massive gas discs.

For completeness, we also decided to establish the domain of validity of the Spitzer model for gas.
Accordingly we depicted in Fig. \ref{fig: numeric vs spitzer} the numeric solution of Eq. \ref{Eq: reduced eq. SG} and the Spitzer solution for different gas Toomre's parameters ranging from 0.001 to 1.
In this plot we considered only the gas profile. 
We observe that for $Q_g \gtrsim 0.1$ the Spitzer model doesn't describe correctly anymore the solution of the hydrostatic equilibrium of a self-gravitating gas disc.
This encouraged us to clarify that a disc in a strong SG regime is a disc where the Toomre's parameter lies below 0.1 and that a value of $Q_g \simeq 1$ indicates a scenario where the gravitational contributions of the star and the gas disc are comparable, meaning neither is significantly greater than the other.
Therefore, the Spitzer profile is valid only when $Q_g \lesssim 0.1$, and consequently, our solution including dust (Eq. \ref{Eq: No const. stopping time, exact high gas mass}) is also valid under this condition.

It is important to note that the standard definition of Toomre's parameter, as introduced by \citep{1964_toomre}, is derived from the analysis of in-plane perturbations in an infinitely thin disc. 
This definition is given by $Q_g=\kappa c_g / (\pi G \Sigma_g)$, where the epicyclic frequency $\kappa$ accounts for Keplerian rotation, the radial pressure gradient, and the radial component of the self-gravity force. 
In contrast, our study disregards in-plane effects and focuses solely on vertical effects. 
Consequently, in our framework, the epicyclic frequency is simply equal to the Keplerian frequency, leading to the definition used in this work: $Q_g=\Omega_K c_g / (\pi G \Sigma_g)$.

\subsection{An approximated solution for any gas disc mass}
\label{subsec: An approximated solution for any gas disc mass}

The previous model can be modified to include the force contribution of the star.
Currently, the vertical stratification of a gaseous disc subject to both the vertical component of the star's and it's own gravity remains unsolved.
However, we know an approximate solution that made use of a biased Gaussian stratification assumption, incorporating all information about SG into a modified scale height \citep{1999_Bertin_Lodato,lodato_2007, 2016_Kratter_Lodato}. 
We will refer to this approximate solution as the BL model.
This model permits a smooth transition between a Keplerian and a self-gravitating gaseous disc by readjusting the scale height with the Toomre's parameter \footnote{We added a factor $\sqrt{2/\pi}$ so as to recover the pressure scale height for a vanishing SG.}: 
\begin{equation}\label{Eq: length scale Lodato}
H_g^{sg} = \displaystyle\sqrt{{2}/{\pi}} \, f(Q_g) \, H_g
\end{equation}
where $f(x)=\frac{\pi}{4 x} \left[ \sqrt{1+\frac{8 x^2}{\pi}} -1 \right]$.
In appendix \ref{app: Bertin-Lodato approximation} we provide a justification of their ansatz.
Accounting for this result, the stratification of a disc where the vertical gravitational influence is primarily influenced by the contributions of gas and the star is:
\begin{equation}\label{Eq: No const. stopping time, Bertin, Lodato stratification}
\left\{
\begin{array}{ll}
\rho_g(\vr,z) & = \displaystyle \frac{\Sigma_g}{\sqrt{2\pi} H_g^{sg}} \exp\left(-\frac{1}{2} \left({z}/{H_g^{sg}}\right)^2\right) \\
\rho_d(\vr,z) & = \displaystyle \frac{\Sigma_d}{\sqrt{2 \pi} \exp\left(\xi^2\right) I_1(\xi^2) H_g^{sg}} \exp\left(-\frac{1}{2} \left({z}/{H_g^{sg}}\right)^2\right) \\ [12pt]
              &   \qquad  \exp\left[ - \xi^2 \left( \exp\left(\frac{1}{2} \left({z}/{H_g^{sg}}\right)^2\right) -1 \right) \right] 
\end{array}
\right.
\end{equation}
The detailed derivation of the integration constant $A(r,\varphi)$ can be found in appendix \ref{app: non constant stopping time, star + gas disc contribution}.
Note that Eq. \ref{Eq: No const. stopping time, Bertin, Lodato stratification} still assumes $\rho_d \ll \rho_g$.
This last vertical profile of gas and dust should be valid from Keplerian discs to massive gas discs (Eq. \ref{Eq: No const. stopping time, exact high gas mass}), that we intend to check next.

\begin{figure}
\centering
\includegraphics[width=\hsize]{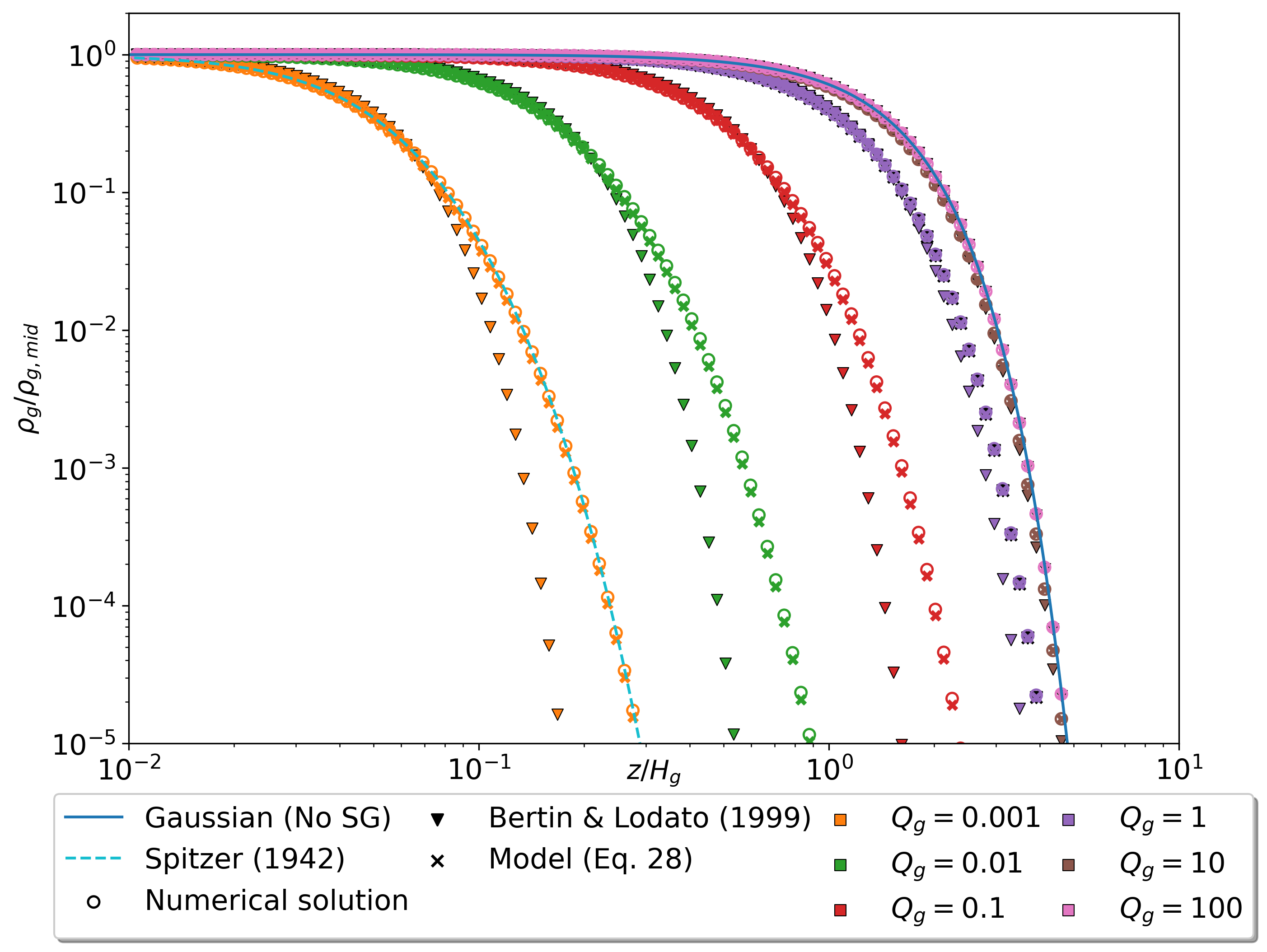}

\caption{Vertical profile of gas for different Toomre's parameters when the dust component is disregarded.
Specifically, we compare the numerical solution of Eq. \ref{Eq: reduced eq. SG} with the BL model and our model Eq. \ref{Eq: stratification single fluid}.}
\label{fig: Gas profile, dust mass negligible}
\end{figure}

\begin{figure}
\centering
\includegraphics[width=\hsize]{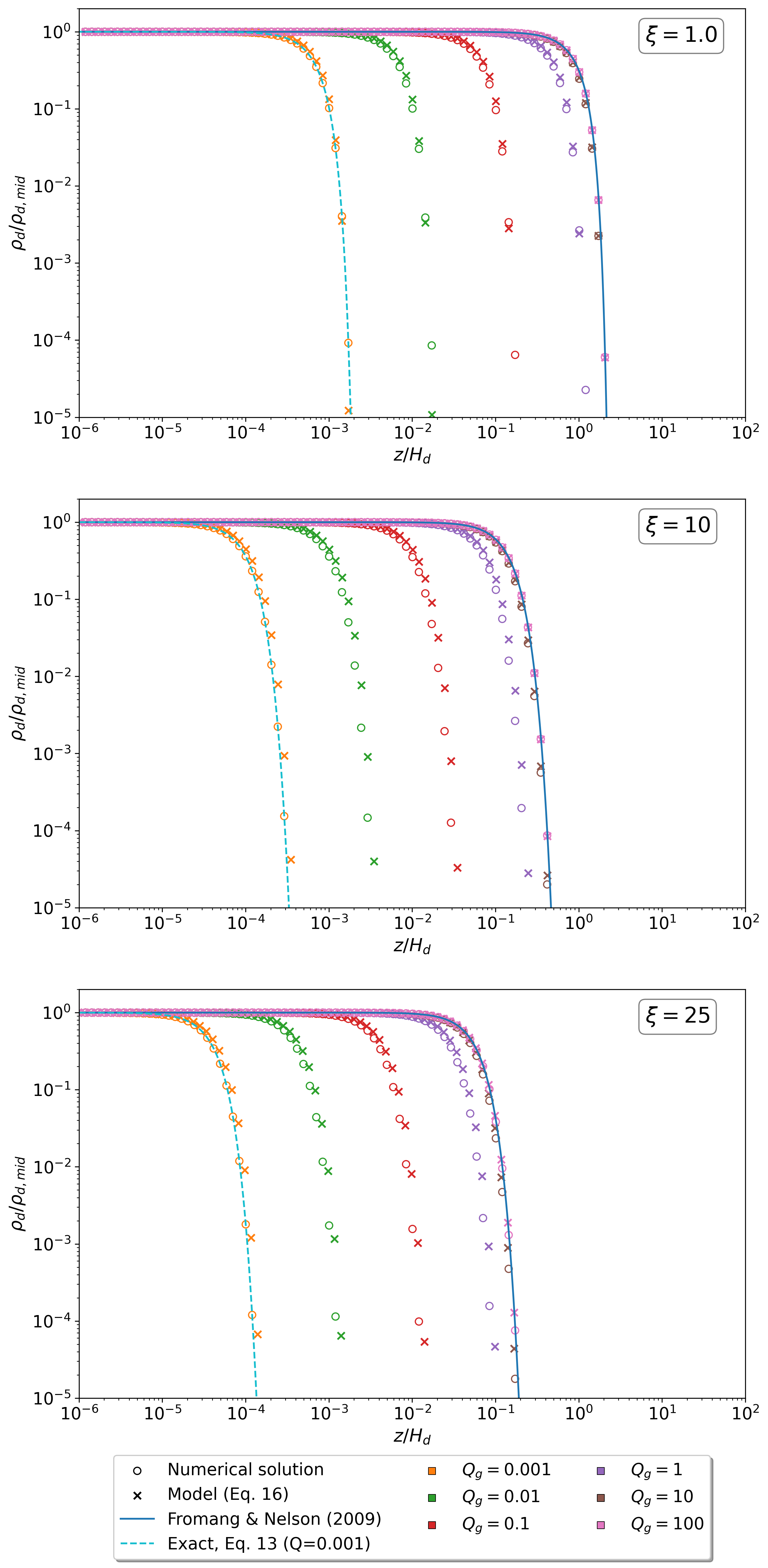}

\caption{Vertical profile of dust density for different gas Toomre's parameters, $Q_g$, and relative dust temperature, $\xi$. 
Specifically, we compare the numerical solution with our model Eq. \ref{Eq: No const. stopping time, Bertin, Lodato stratification}.
Additionally, we added the limiting cases of Fromang \& Nelson profile, valid in absence of SG, and the exact solution that we derived (Eq. \ref{Eq: No const. stopping time, exact high gas mass}), valid for massive gas discs.
} 
\label{fig: gas and dust vertical density profile stopping time with vertical profile}
\end{figure}

First, we aim to establish the validity domain of the BL profile for gas.
In Fig. \ref{fig: Gas profile, dust mass negligible} we depicted the numeric solution of Eq. \ref{Eq: reduced eq. SG} and compared it with the BL model for different gas Toomre's parameters ranging from 0.001 to 100.
Additionally, we added the limiting cases of the Spitzer profile (for $Q_g=0.001$) and the classic Gaussian profile valid in absence of SG ($Q_g=\infty$).
We also added the model that we will present in Sect. \ref{subsec: the approximated solution}, but we will not comment on it for now.
We observe that the biased Gaussian model fits accurately the numerical solutions for $Q_g\geq 1$ across all $z$ values.
However, for lower Toomre's parameter values, $Q_g\leq 1$, discrepancies between the numerical solution and the biased Gaussian model become apparent when $z/H_g \gtrsim 4 \sqrt{Q_g}$.
This could be explained twofold: first a Gaussian is used for approximating a hyperbolic function, a solution that holds true close to the midplane but incorrect at intermediate $z/H_g$, second even if the  BL model is Gaussian at infinity, it nevertheless has the wrong scale height coefficient.
We will propose a discussion about this issue in Sect. \ref{sec: Bertin-Lodato and Spitzer models: a discussion}.
In summary the BL model accurately bridge the gap between light discs SG ($Q_g=\infty$) and discs of moderate SG ($Q_g\simeq 1$), but shows significant deviations for massive discs ($Q_g\leq 1$).

Additionally, the vertical profile of dust corresponding to the aforementioned gas profile is displayed in the three panels of Fig. \ref{fig: gas and dust vertical density profile stopping time with vertical profile} for three distinct relative gas to dust temperatures: $\xi \in [1, 10, 25]$.
Similar to the previous case, we present the numerical solution of Eq. \ref{Eq: reduced eq. SG} and our proposed model for dust in Eq. \ref{Eq: No const. stopping time, Bertin, Lodato stratification}. 
In the same figures, we have also included the exact stratification of \citet{2002A_takeuchi} and \citet{2009b_fromang_nelson}, which are valid for Keplerian discs with negligible SG. 
Additionally, we depicted our new analytic solution presented in Eq. \ref{Eq: No const. stopping time, exact high gas mass} for massive gas discs.
We observe that for all relative temperatures our dust model asymptotically approaches the analytic models of \citet{2002A_takeuchi} and \citet{2009b_fromang_nelson} for lighter discs and our exact solution for more massive gas discs. 
Further, similar to the previous case, we find that noticeable deviations between the numerical and dust model appear for increasing relative temperatures, low gas Toomre parameters and ${z/H_d \gtrsim \xi Q_g}$, but these discrepancies are not as pronounced as they are with gas.

\subsection{Spitzer and Bertin-Lodato models: a discussion on boundary conditions}
\label{sec: Bertin-Lodato and Spitzer models: a discussion}

For this Sect. we are concerned by the hydrostatic equilibrium of gas in absence of dust:
\begin{equation}\label{Eq: constant stopping time, SG of gas}
\displaystyle \Tcg^2 \partial_{zz} \ln{\left(\rho_g\right)}  
=  - \Omega_K^2  - {4\pi G}  \rho_g  \\
\end{equation}
As for any differential equation, this hydrostatic equilibrium is necessarily completed by boundary conditions satisfied by the scalar field and its derivative, which ensure the uniqueness of the solution.
The Von Neumann condition is straightforward since the density is expected to reach its maximum value at the midplane, which translates to $\partial_z \rho_g (z=0)=0$.
The Dirichlet condition is set equivalently either by a simple boundary condition in the midplane, or by a coherent condition such as the one proposed in Eq. \ref{Eq: column density}.
We called this condition coherent because the surface density will affect the scale-height definition, which in turn affects the surface density.
Further, the condition defined by Eq. \ref{Eq: column density} implicitly implies the integrability of the density, which means that it should vanish at infinity while maintaining approximately a constant value close to the midplane.
In other words, when "seen" from large heights the gas density can be considered as a Dirac delta function at the disc midplane.
Consequently, at large heights the gas stratification is simply governed by the star contribution and the gravity of a thin layer making the density to adopt the classic Gaussian profile with an absolute value correction.
All these clarifications can be summarised as:
\begin{equation}\label{Eq: constraints SG of gas}
\left\{
\begin{array}{llll}
\rho_g(z=0) &=& \rho_{g,{\rm mid}} & \text{or} \quad \text{Eq. \ref{Eq: column density}} \\
\lim\limits_{z\rightarrow \infty} \rho_g(z) & \simeq &  \rho_{g,{\rm mid}} \exp\left( -\frac{1}{2} \left( \frac{z}{H_g} \right)^2 - \frac{2 |z|}{Q_g H_g} \right) & \\
\partial_z \rho_g (z=0) & = & 0 &
\end{array}
\right.
\end{equation}
The model of BL captures correctly the gravity terms close to the midplane (up to the second order) thanks to a constant, modified scale height, which in turns forbids the asymptotic connection with above asymptotic profile.  
Nevertheless, this model informs us about the length-scale on which the gravity of the star and the disc prevail, before connecting back to the Keplerian disc.
In other words, $H_g^{sg}=\sqrt{2/\pi} f(Q_g) H_g$ is the length where the gravity of the disc cannot be disregarded: in the far-zone regime, after few $H_g^{sg}$, the disc can be thought as razor-thin and the standard Gaussian profile, augmented by the $|z|$ correction, should be retrieved.
The Spitzer model ignores these considerations because it assumes that the stellar term is zero when the Toomre parameter is very low. 
However, we have two objections: first, the star's gravity must be present for large heights, more precisely beyond $\sim H_g/Q_g$. 
Second, the Spitzer model can only disregard the star's gravity for very low $Q_g$ values, below $\sim 0.1$, at which the disc is expected to be highly unstable, making the assumption of hydrostatic equilibrium invalid.

We would like to clarify that we are not questioning the models of Spitzer and BL, which remain valid within the framework of their hypotheses. 
Namely, an approximate solution near the midplane.
However, this detailed analysis on the boundary conditions has enabled us to reconsider the issue of stratification and refine the constraints, ultimately leading us to discover very precise and more general approximate solutions.
We will present these solutions in Sect. \ref{sec: constant stopping time and contribution of dust}.

\subsection{Summary}

Our approximate model for gas and dust matches correctly the numerical and analytic solutions for $Q_g \gtrsim 1$.
For smaller Toomre parameters the Gaussian model (Eq. \ref{Eq: No const. stopping time, Bertin, Lodato stratification}) tends towards the analytic solution that we found (Eq. \ref{Eq: No const. stopping time, exact high gas mass}) but it deviates from ${z/H_g \gtrsim 4 \sqrt{Q_g}}$, or ${z/H_d\gtrsim 3 Q_g}$, for gas and dust profiles, respectively.
However this is of little importance since at these heights the density of gas and dust can be considered negligible compared to the midplane values.

The stratifications found in this Sect. are valid only in the limit where the gravitational contribution of dust can be disregarded.
However, this assumption might not always hold true, considering that the observed low quantity of dust mass in Class II discs might not adequately account for the mass of discovered exoplanets \citep{2016_ansdell, 2018_manara, 2021_mulders}.
This disagreement may imply the presence of more dust in discs than initially anticipated.
Further, it is often considered that dust SG should be disregarded because of the low dust-to-gas column density ratio, $Z=\Sigma_d/\Sigma_g$, inferred from the ISM.
Nonetheless, we firmly believe that the appropriate measure for quantifying dust SG in the vertical orientation should be the dust-to-gas volume density ratio,  $\rho_d/\rho_g \simeq \frac{\Sigma_d}{\Sigma_g} \frac{H_g}{H_d}$. 
This quantity could be substantial for highly settled dust, a common occurrence frequently observed in PPDs \citep{2016_pinte,2020_villenave, 2022_villenave}.
Finally, disc masses derived from (sub-)mm fluxes are underestimated compared to SED-inferred masses \citep{2019_ballering, 2020_ribas}.
This is further strengthened by the work of \citet{2024_vorobyov}, who calculated the synthetic dust mass of a known numerical model.
They found that the inferred mass from the observational estimates was two orders of magnitude smaller than the actual value, supporting the idea of a hidden dust reservoir within the disc. 
All aspects raised in this paragraph reinforce the importance of considering dust SG, prompting us to incorporate it into our analysis in the following section with a simpler assumption on the stopping time.

\section{Constant stopping time and gravitational contribution of dust}
\label{sec: constant stopping time and contribution of dust}

The constraint of a vertical dependence in the stopping time is very restrictive and complicates the resolution of the gas hydrostatic equation and transport equation for dust.
Thus, in this Sect. we use the simple assumption $\tau_f = \tau_{f,{\rm mid}}$, which permits to rearrange the hydrostatic equations in a symmetric way where both gas sound speed and dust diffusivity are constant:
\begin{equation}\label{Eq: constant stopping time, SG of gas and dust}
\left\{
\begin{array}{cc}
\displaystyle \Tcg^2 \partial_{zz} \ln{\left(\rho_g\right)}  &= \displaystyle - \Omega_K^2  - {4\pi G} \left( \rho_g + \rho_d \right) \\ [8pt]
\displaystyle \Tcd^2 \partial_{zz} \ln{\left(\rho_d\right)}  &= \displaystyle - \Omega_K^2  - {4 \pi G} \left( \rho_g + \rho_d \right)
\end{array}
\right.
\end{equation}
Here, $\Tcd=  {\xi}/{\sqrt{\xi^2+1}} \, c_{d, {\rm mid}}$ is an effective dust sound speed.

\subsection{The approximated solution}
\label{subsec: the approximated solution}

We would like to emphasise that the primary challenge in solving Eqs. \ref{Eq: constant stopping time, SG of gas and dust} lies in the coupled non-linear nature of the differential equations. 
More importantly, the simultaneous management of different scales presents a significant obstacle. 
For instance, considering the gas alone, we can distinguish hastily at least three possible height scales: the pressure scale height ($H_g$), the scale height of the influence of the gas mass ($\sim Q_g H_g$), and the scale height of the influence of dust mass ($\sim Q_d H_g$).
For the latter, we further anticipate a weighting factor that depends on the relative temperature, $\Txi={\Tilde{c}_g}/{\Tilde{c}_{d,{\rm mid}}}$, which is not trivial at first sight. 
The presence of these different scale heights makes the resolution particularly difficult and precludes for bi-fluids the use of the method employed by \citet{1999_Bertin_Lodato}, as detailed in Appendix \ref{app: Bertin-Lodato approximation}.
We highlight that in this work we use 
\begin{equation}
Q_d=\frac{\Tcd \Omega_K}{\pi G \Sigma_d}=\sqrt{ \frac{\az}{\az+(1+\az)\,\sto} }\;\frac{\Sigma_g}{\Sigma_d} Q_g\,,
\end{equation}
which aligns with the definition of \citet{2021_klahr_schreiber} in the limit of small vertical diffusivity compared to the Stokes number.

As in previous Sect., the set of Eqs. \ref{Eq: constant stopping time, SG of gas and dust} can be ultimately reduced to a single equation with a transformation that permits to retrieve the dust profile:
\begin{equation}\label{Eq: unique Liouville eq}
\left\{
\begin{array}{lll}
\Tcg^2 \partial_{zz} \ln\,\left(\displaystyle \frac{\rho_g}{\rho_{g,{\rm mid}}}\right) 
          &=& -\Omega_K^2 \\
          & & \displaystyle- 4 \pi G \left[ \rho_g + 
                               \rho_{d,{\rm mid}} \, \left( \frac{\rho_g}{\rho_{g,{\rm mid}}} \right)^{\Txi^2}  \right] \\
\displaystyle \rho_d = \rho_{d, {\rm mid}} \left( \frac{\rho_g}{\rho_{g,{\rm mid}}} \right)^{\Txi^2} & &
\end{array}
\right.
\end{equation}
where $\rho_{i,{\rm mid}}$ is the midplane density of phase $i$.
Equation \ref{Eq: unique Liouville eq} is essentially a 1D Liouville equation \citep{1853_liouville} with a constant term and two exponential nonlinearities.
Even when the star's gravity is negligible, $\Omega_K=0$, solving Eq. \ref{Eq: unique Liouville eq} is a challenging endeavour that represents an active field of research in mathematical physics \citep{2018_mancas}.
To our knowledge, there are very few exact solutions available, but unfortunately, they do not apply to our specific problem. 
Furthermore, we are also interested in the smooth transition between a Keplerian disc into a self-gravitating disc dominated either by gas, or by dust or by both phases.
This implies that we cannot neglect any term in Eq. \ref{Eq: unique Liouville eq}.
However, motivated by the underlying structure of new analytical solutions to Eq. \ref{Eq: constant stopping time, SG of gas and dust} in specific cases (see Appendix \ref{app: constant stopping time}) and the method that we developed in Appendix \ref{app: A general approximated solution}, we managed to build an accurate approximate solution that is valid for all relative temperatures and in all regimes of SG, from Keplerian discs to massive discs made of gas and/or dust and that respect the physical boundary conditions mentioned in Sect. \ref{sec: Bertin-Lodato and Spitzer models: a discussion}.
This new approximated solution is:
\begin{equation}\label{Eq: stratification gas and dust}
\left\{
\begin{array}{ll}
\displaystyle 
\rho_g(\vr, z) = \rho_{g,{\rm mid}} \exp\left[
-\frac{1}{2} \left(\frac{z}{H_g} \right)^2 \right.
            & \displaystyle - \sqrt{\frac{\pi}{2}} \frac{1}{Q_g^{sg}} \Tf{}{\frac{z}{H_g^{sg}}} \\ [10pt]
            & \displaystyle \left. - \sqrt{\frac{\pi}{2}} \frac{1}{\Txi^2 Q_d^{sg}} \Tf{}{\frac{z}{H_d^{sg}}}  \right] \\
            \displaystyle
\rho_d(\vr, z) = \rho_{d,{\rm mid}} \left[ \rho_g/\rho_{g,{\rm mid}} \right]^{\Txi^2}
\end{array}
\right.
\end{equation}
with:
\begin{equation}\label{Eq: parameters stratification gas and dust}
\left\{
\begin{array}{lll}
\displaystyle \Tf{}{z} &=& \displaystyle \exp\left(-\frac{1}{2} z^2 \right) 
                       + \sqrt{\frac{\pi}{2}} \, z \, \text{erf}\left(\frac{z}{\sqrt{2}}\right) - 1  \\
k_1 &=& 1 \, \Big/ \sqrt{1+ \sqrt{\frac{\pi}{2}} \left(\frac{1}{Q_g^{3D}} + \frac{1}{Q_d^{3D}} \right)} \\
H_i^{sg} &=& k_1 H_i \\
Q_{i}^{3D}      &=& \displaystyle \sqrt{ \frac{\pi}{2} } \frac{\Omega_{K}^2}{\pi^2 G \rho_{i,{\rm mid}}} \\ [8pt]
Q_i^{sg} &=& \displaystyle Q_i^{3D} \left({\Omega_{sg}}/{\Omega_K}\right)^2 \\
\Omega_{sg} &=& \Omega_K/k_1
\end{array}
\right.
\end{equation}
where $\Txi = {\Tilde{c}_g}/{\Tilde{c}_{d,{\rm mid}}}$, `erf' is the error function, and the function $f$ is the same as the one obtained with BL's method.
We emphasize that our definition of the 3D Toomre parameter, $Q_i^{3D}$, slightly differs from the one proposed by \citet{2010_mamatsashvili}, that is, by a factor of approximately 1.6.
However, we believe our definition is correct because, for negligible SG in the vertical direction, it aligns with the 2D Toomre parameter.
Without going into further detail, we see that our problem is governed by three height scales, the interpretation of which will be provided in Section \ref{subsec: physical interpretation}.
It is curious that the relative temperature $\Txi$ does not affect explicitly any physical quantity establishing the settling, like $H_i^{sg}$ or $Q_i^{3D}$.
However, it could be indirectly present in the $\rho_{i,{\rm mid}}$ definition.
It seems that its unique role is to act as a conversion parameter between the gas and dust layering.
Finally, in a fairly convenient manner, the $\Theta$ function satisfies next relation at infinity:
\begin{equation}
\lim\limits_{z\rightarrow \pm \infty} \Theta\left(\frac{z}{H_g^{sg}}\right) = \sqrt{\frac{\pi}{2}} \frac{|z|}{H_g^{sg}} 
\end{equation}
which makes $\rho_g$ to adopt the asymptotic behaviour expected from Sect. \ref{sec: Bertin-Lodato and Spitzer models: a discussion} discussion.

\subsection{Validation}

\begin{figure*}[!th!p]
\centering
\resizebox{0.85\hsize}{!}
{
    \includegraphics[width=\hsize]{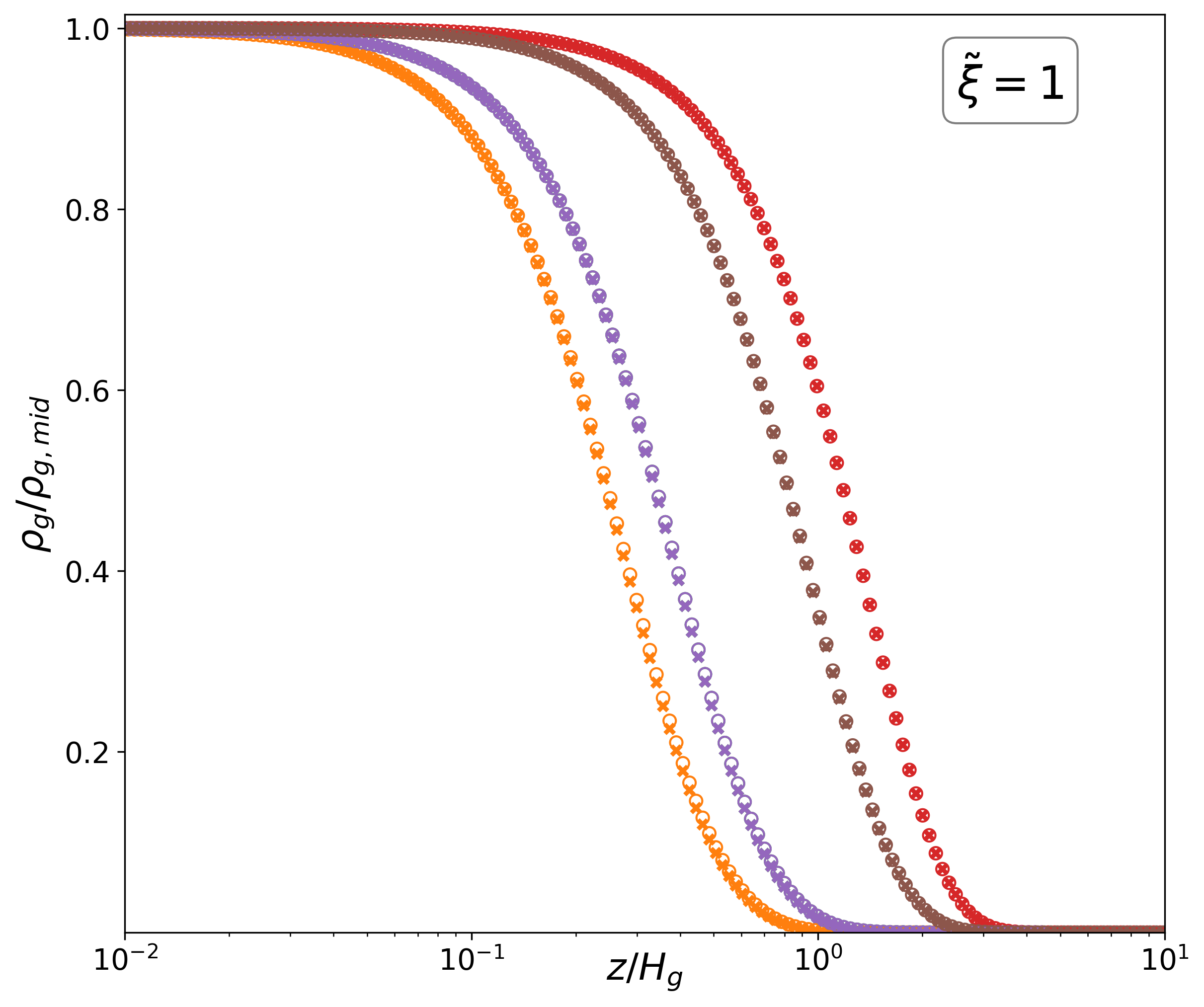}
    \includegraphics[width=0.90\hsize]{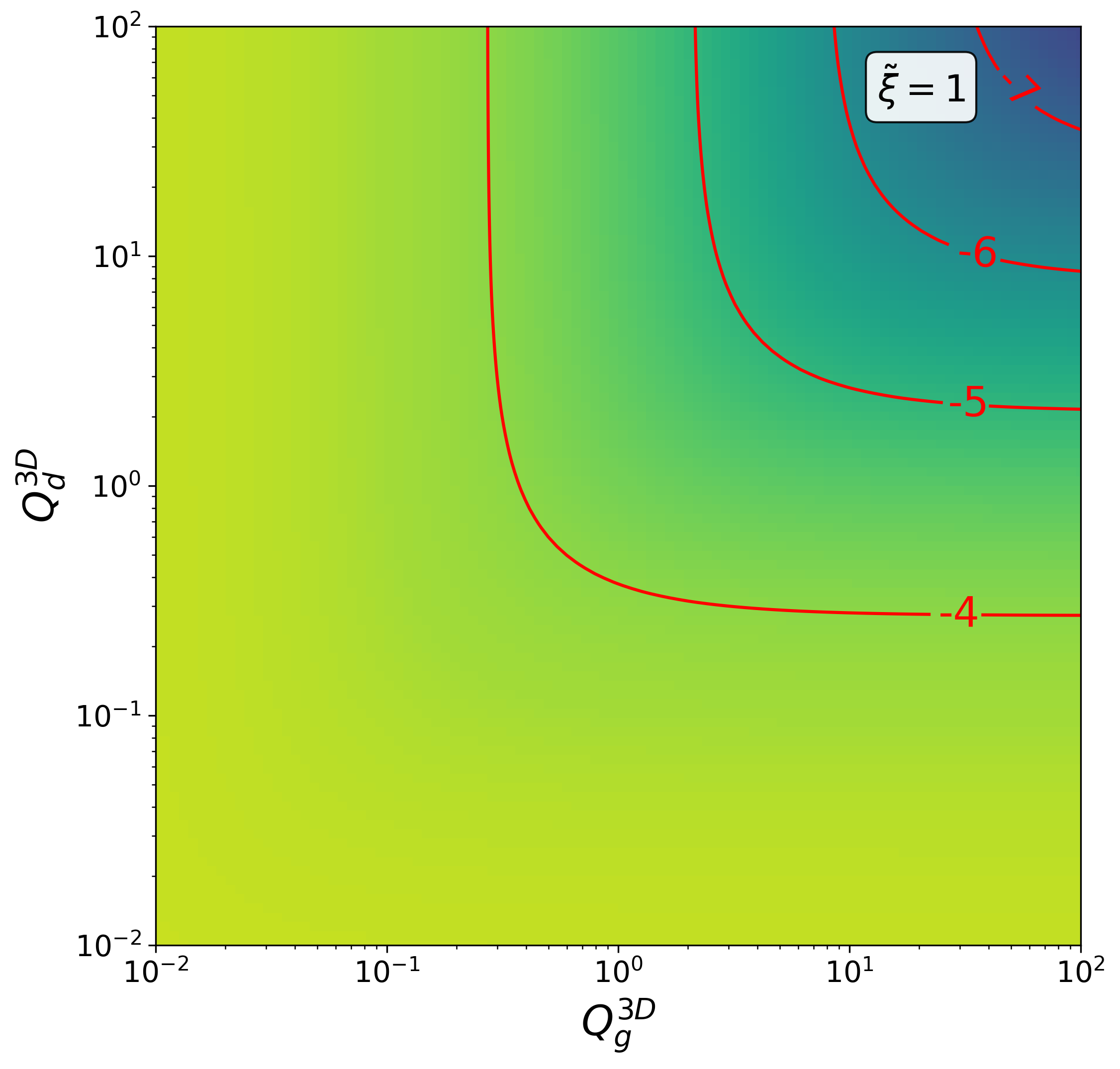}
}
\vfill
\resizebox{0.85\hsize}{!}
{
    \includegraphics[width=\hsize]{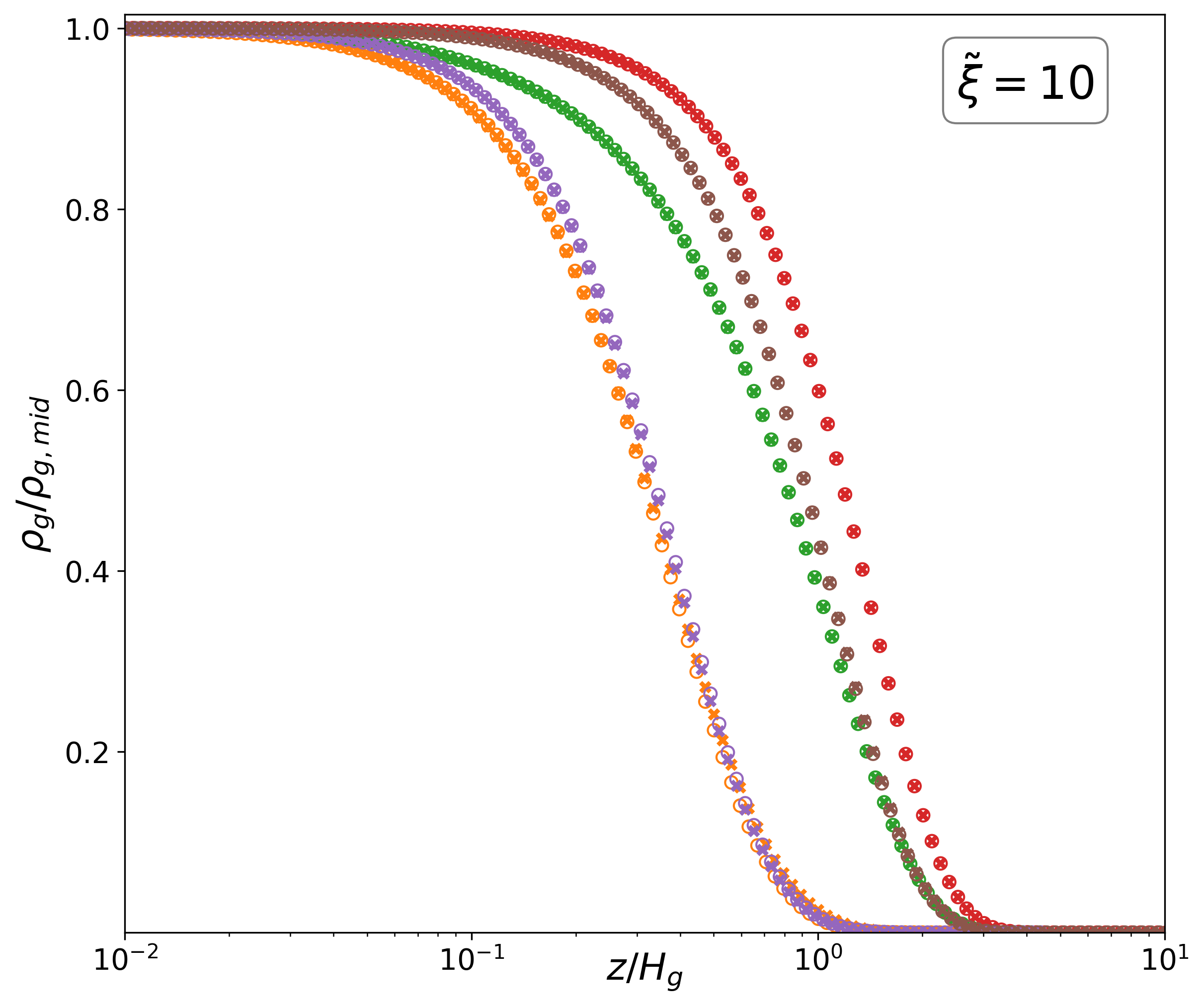}
    \includegraphics[width=0.9\hsize]{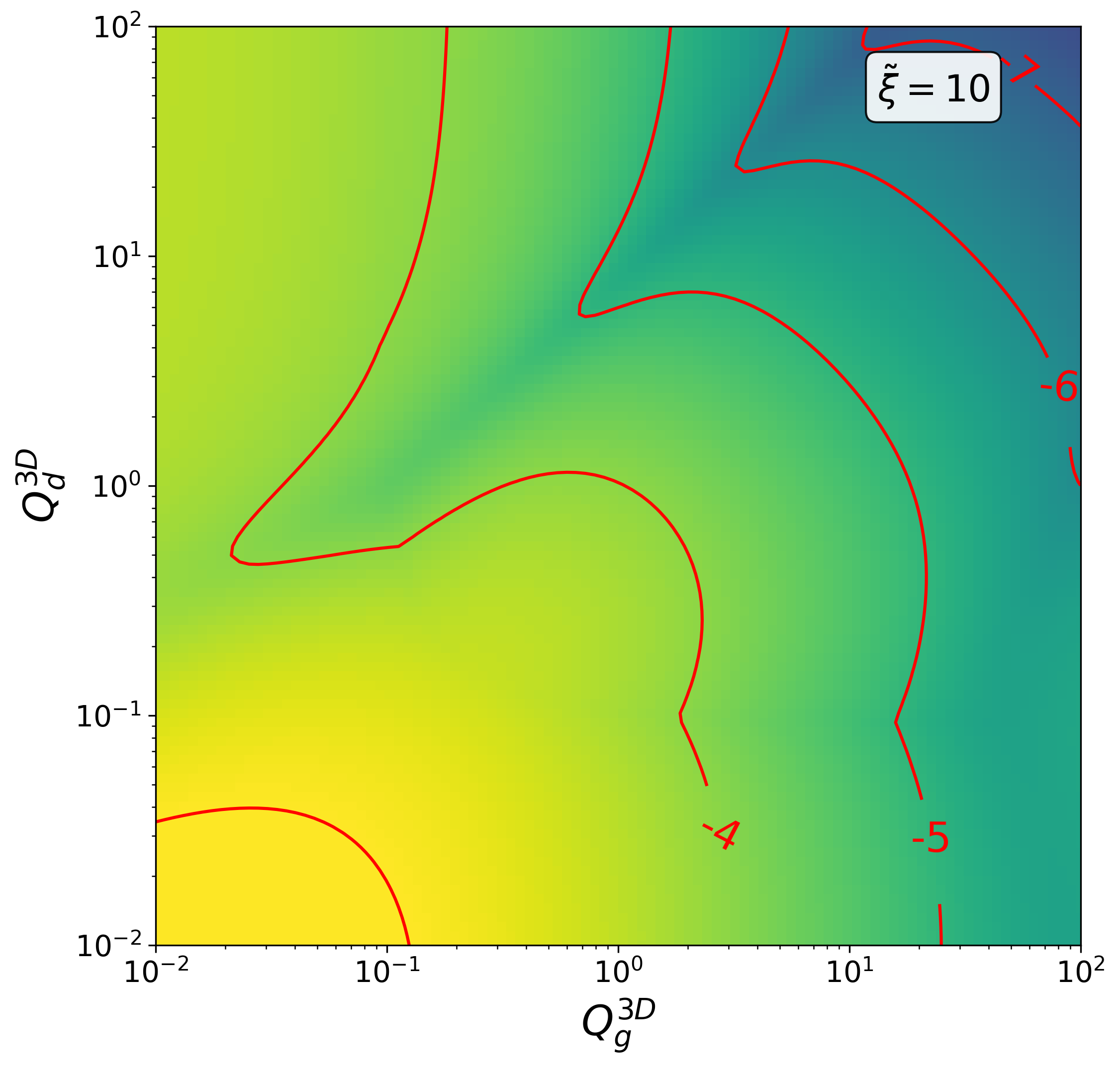}
}
\vfill
\resizebox{0.85\hsize}{!}
{
    \includegraphics[width=\hsize]{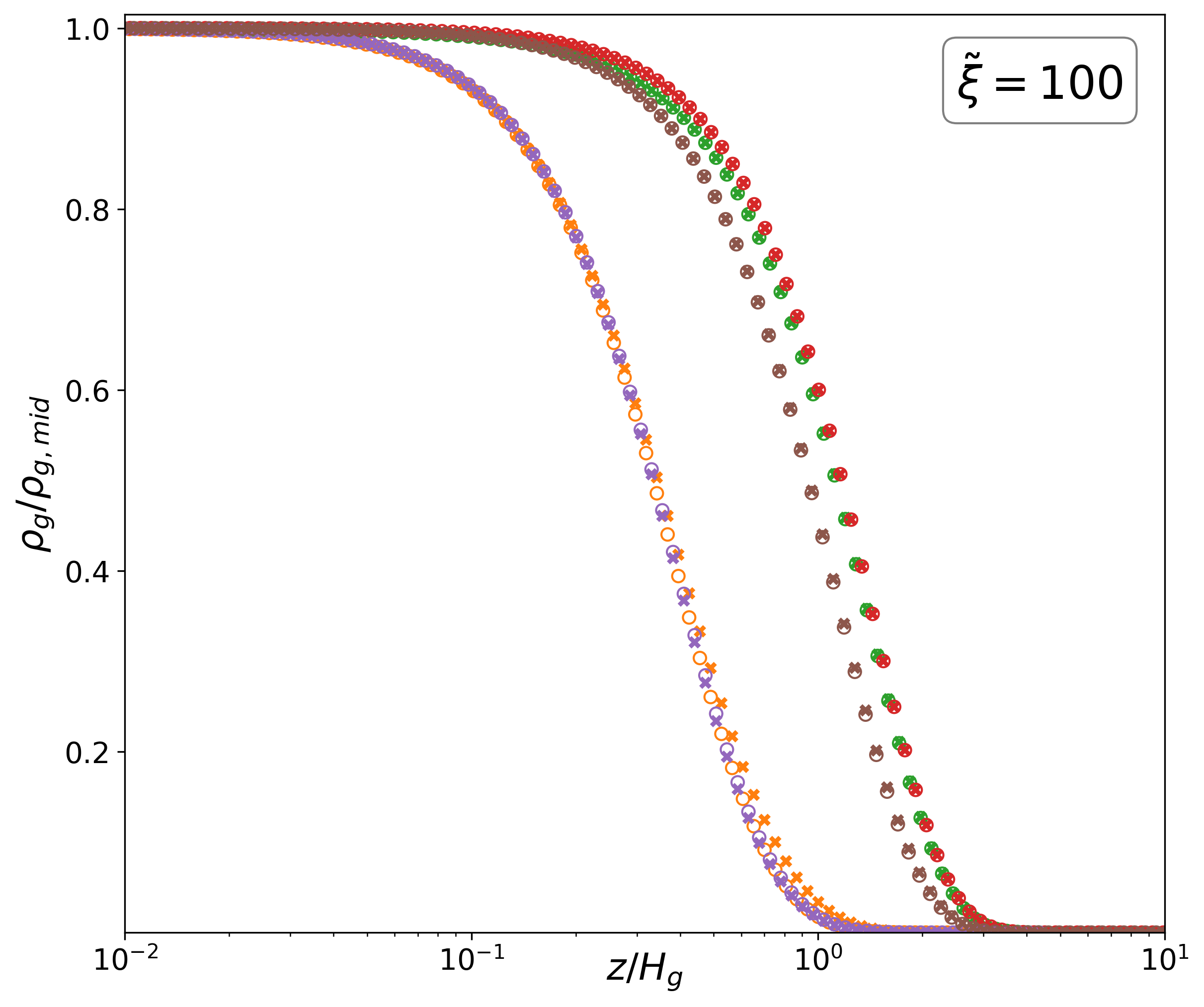}
    \includegraphics[width=0.9\hsize]{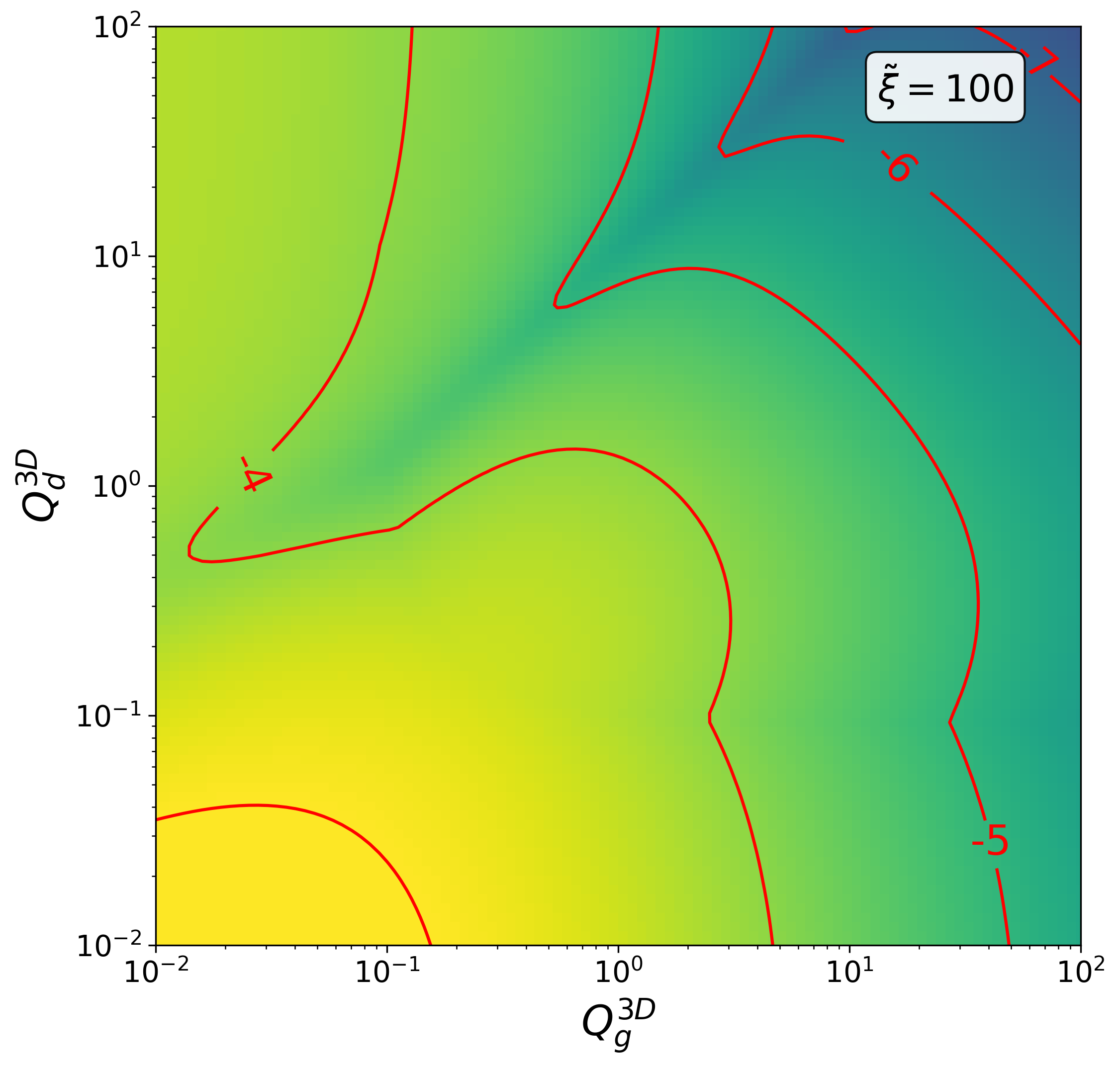}
}
\vfill
\resizebox{0.85\hsize}{!}
{
    \includegraphics[width=\hsize]{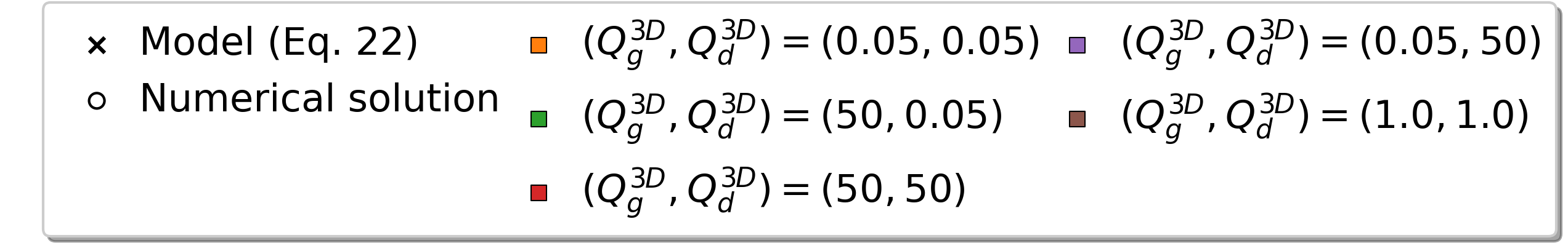}
    \includegraphics[width=0.9\hsize]{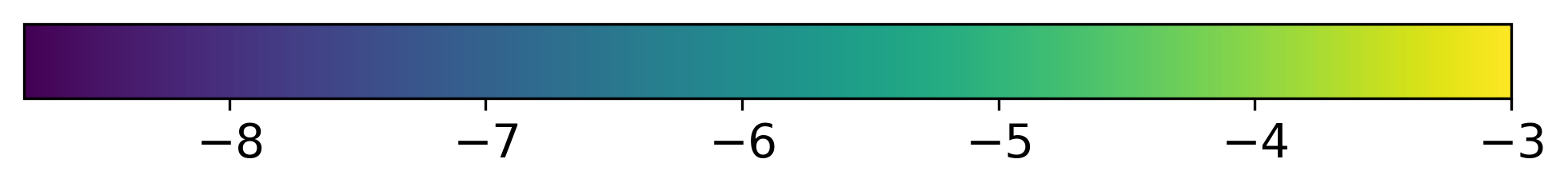}
}
\vfill
\caption{Comparison between the numerical solution of Eq. \ref{Eq: constant stopping time, SG of gas and dust} and our approximated solution (Eq. \ref{Eq: stratification gas and dust}) for different Toomre parameters of gas ($Q_g^{3D}$), dust ($Q_d^{3D}$) and relative effective gas-to-dust temperature ($\Txi$). \\
Left column: Vertical profile of gas density. Right column: L2 norm error map in $\log_{10}$ scale.
} 
\label{fig: Gas profile with dust mass and L2 norm}
\end{figure*}

For validating the accuracy of our solution, there are two techniques. 
The first involves comparing our results with analytic and approximate solutions in limiting cases, such as those presented in Appendices \ref{app: Bertin-Lodato approximation} and \ref{app: constant stopping time}. 
While this validation is left to the reader, we would like to confirm that we have verified that, close to the midplane, our solution yields the scale heights associated with all limiting cases.
Particularly, when dust is absent, a Taylor expansion up to second order in function $\Theta$ in Eq. \ref{Eq: stratification gas and dust} permits to retrieve BL model.
The second validation method is based on numerical comparison, which we will discuss next.

Now, we will estimate the accuracy of our self-gravitating bi-fluid model.
To do so, we quantified the deviation between the numerical solution and our model by estimating the following L$_2$ norm:
\begin{equation}
||\varepsilon||_2 = \frac{1}{N_z} \left(  \sum\limits_{i=0}^{N_z} \left|\frac{\rho_{g,\text{model}}(z_i)}{ \rho_{g,\text{num}}(z_i) }-1\right|^{2} \right)^{1/2}
\end{equation}
where $N_z$ is the number of points.
In Fig. \ref{fig: Gas profile with dust mass and L2 norm} we compared the numerical solution of Eq. \ref{Eq: constant stopping time, SG of gas and dust} with our approximated solution (Eq. \ref{Eq: stratification gas and dust}) for different relative temperatures ($\Txi$) and different gas ($Q_g^{3D}$) and dust ($Q_d^{3D}$) Toomre parameters.
Specifically, in the left column we plotted the vertical profiles of gas and, in the right column, we present a map of above L$_2$ norm for gas and dust Toomre parameters, ranging from 0.01 to 100, and a fixed relative temperature. 
Specifically, for each triplet $(Q_g^{3D}, Q_d^{3D}, \Txi)$, we employed $N_z=50 000$ points spaced evenly in a log scale in the range $z/H_g^{sg} \in [1/(100 \Txi), 5]$ \footnote{We added the 0 point as well}. 
Using a log scale enables us to capture both small and large scale variations associated with dust and gas. 
We found that the relative deviation is on the order of $10^{-4}$ and $10^{-3}$ for all our parameters. 
Such low deviations confirm that our approximated solution is a generally correct description for self-gravitating discs composed of gas and dust.

Now, we provide a physical interpretation of all quantities involved in the gas-dust layering that we just found.

\subsection{Physical interpretation}
\label{subsec: physical interpretation}

\begin{figure}
\centering
\includegraphics[width=\hsize]{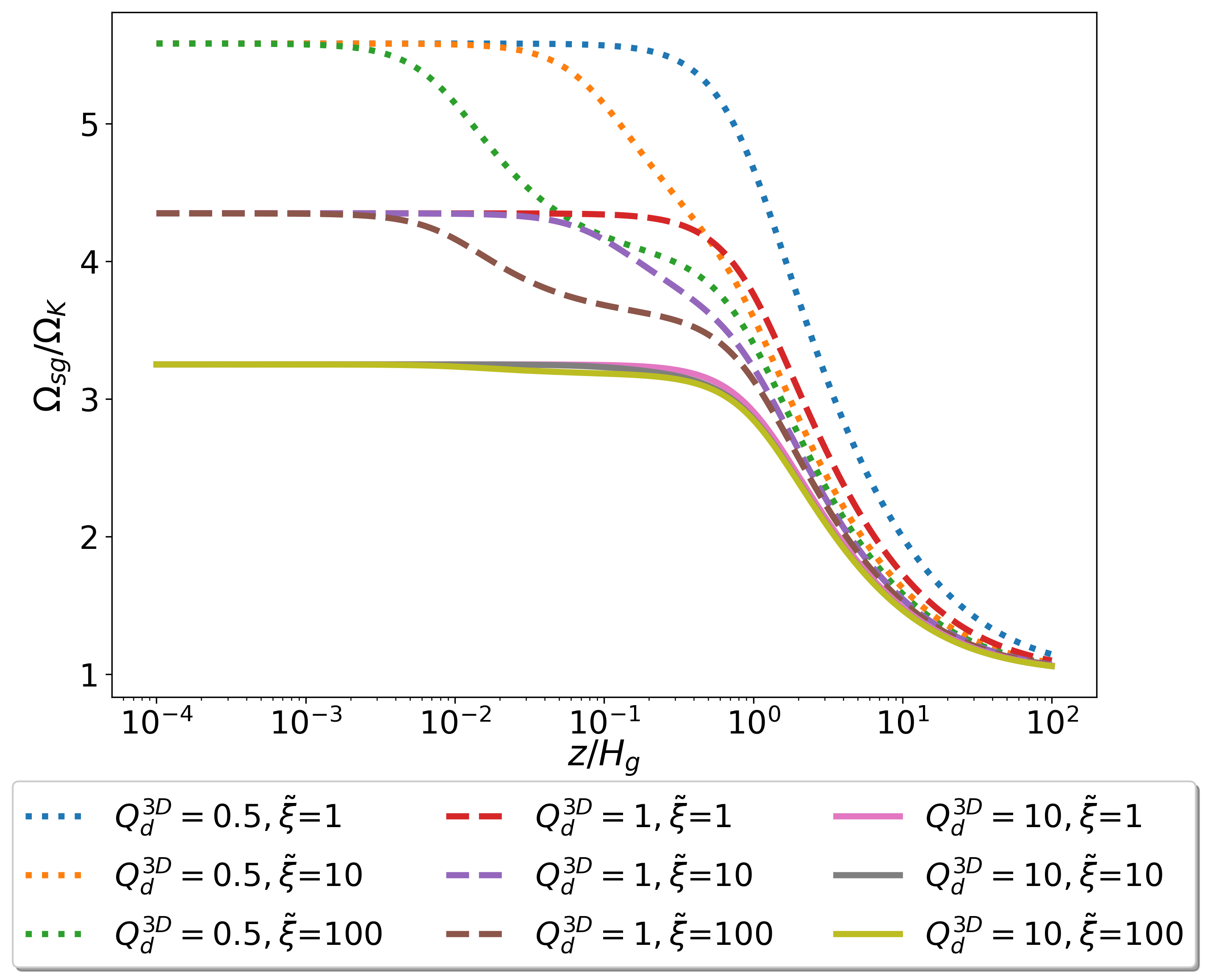}

\caption{Self-gravitating differential rotation in a marginally gas-stable disc ($Q_g^{3D}=0.5$) for various dust Toomre parameters ($Q_d^{3D}$) and relative temperatures ($\Txi$).}
\label{fig: omega SG}
\end{figure}

We now provide a physical interpretation of the quantities involved in the stratification described by Eqs. \ref{Eq: stratification gas and dust}-\ref{Eq: parameters stratification gas and dust}. 
In this paragraph we will primarily focus on the gas length scales, as the dust length scales can be directly obtained by multiplying by $\Txi$. 
The first noticeable scale is the standard pressure scale height, $H_g$, whose perceptible role is to account for the star gravity and to allow for a connection with the background Gaussian layering.
In the presence of both phases SG, $H_g^{sg}$ and $H_d^{sg}$ represent the lengths over which the gravity generated by the gas and dust profiles act, respectively. 
In other words, these are the widths of the gas and dust layers.
Specifically, in absence of dust, $H_g^{sg}$ equals the length scale of the BL biased Gaussian model (see Eq. \ref{Eq: length scale Lodato}). 
These scale height definition involve particularly the quantity $k_1$ provided by Eq. \ref{Eq: parameters stratification gas and dust}, which is constructed as a combination of the root mean square and harmonic average of the Toomre parameters of the gas and dust, which also includes the star contribution (the unity term).
Somehow, this writing shows the competition between all terms for dominating the vertical settling.
More interestingly, the harmonic average of the gas and dust Toomre parameters naturally emerges:
\begin{equation}\label{Eq: Toomre bi-fluid}
Q_{\text{bi-fluid}}^{3D} = \left( \frac{1}{Q_g^{3D}} + \frac{1}{Q_d^{3D}} \right)^{-1}
\end{equation}
It is immediately interpreted as the general Toomre parameter for a bi-fluid, self-gravitating disc when dust is embedded in a turbulent gaseous environment. 
An interesting property of this harmonic average is that when the mass of one of the two phases predominates, the bi-fluid Toomre parameter will adopt its value.
We strongly believe that this definition of the Toomre parameter, which naturally emerges from first principles, is the appropriate quantity for quantifying self-gravity in bi-fluid systems. 

We propose another axis of analysis based on differential rotation. 
If all terms inside the exponential of Eq. \ref{Eq: stratification gas and dust} are factored by $z^2/2$, we can rewrite the gas density profile as a simple Gaussian, with a scale height defined by $H_g^{sg}(z)=c_g/\Omega_{sg}(z)$, where $\Omega_{sg}$ acts as a modified differential rotation due to SG. 
Indeed, everything happens as if the disc was more massive close to the midplane.
The expression for the modified rotation due to SG is:
\begin{equation}\label{Eq: sg rotation profile}
\frac{\Omega_{sg}(z)}{\Omega_K} =
\sqrt{1 + \sqrt{2\pi}  \left( \frac{H_g}{z} \right)^2 \left[ \frac{1}{Q_g^{sg}}  \Tf{}{\frac{z}{H_g^{sg}}} + \frac{1}{\Txi^2 Q_d^{sg}}  \Tf{}{\frac{z}{H_d^{sg}}} \right] }
\end{equation}
It is important to note that SG influences the dynamics of massive discs, causing significant deviations from the Keplerian rotation profile \citep{Lodato_2023, 2024_veronesi, 2024_speedie}.
However, in our framework, the above differential rotation is a theoretical construct that helps us understand the problem from a different perspective, but it is not directly observable nor related to the kinematic signatures of SG.
As explained in Sect. \ref{subsec: An exact solution for massive gas discs}, our study focuses solely on vertical effects. 
To establish a clear connection between the rotation profile and vertical stratification of massive discs, we would need to remove the slab approximation. 
Indeed, this approximation assumes that the self-gravity (SG) of a volume element is influenced only by the mass distribution at that specific location.
Nonetheless, the reduction in scale height due to SG also decreases the pressure support, likely impacting the rotation profile \citep[Eq. 13]{2013_nelson}. 
In Fig. \ref{fig: omega SG} we depicted above self-gravitating frequency for different dust Toomre's parameters and relative temperatures.
For all cases, we fixed $Q_g^{3D}=0.5$, which is the standard value found in 3D simulations where the gravitational instability of gas is self-regulated for slow cooling.
We found subsequent deviations to the Keplerian frequency that range between 2 to 5 times the Keplerian frequency.
In some cases, these deviations can be felt well above the midplane, up to 5 pressure scale heights.
Our plots of the self-gravitating frequency also show that our problem is governed by various scale heights.
This is particularly explicit for the curve $(Q_d^{3D},\Txi)=(0.5, 100)$ where we see that at $z/H_g \simeq 10^{-3}$ and $z/H_g \simeq 10^{-1}$ the slope of the curve suddenly changes.
This behaviour could be mostly seen for high values of the relative temperature.
Finally, at the midplane, $z=0$, the ratio $\Omega_{sg}/\Omega_K$ equals $1/k_1$.
Based on this result, we propose in next paragraph a possible redefinition of the Toomre parameters governing our problem.

In Eq. \ref{Eq: parameters stratification gas and dust}, we introduced self-gravitating Toomre parameters: $Q_i^{sg}$. 
It can be readily shown that these parameters align with the standard definition for light discs, since in this limit we have:  $\Omega_{sg} \xrightarrow[]{} \Omega_K$.
In this context, it is important to recall that the Toomre stability criterion, $Q_g > 1$, indicates the condition under which razor thin discs are stable against perturbations \citep{1964_toomre}.
An intriguing aspect of our definitions is the possibility of a situation where, for instance, $Q_g<1$ for the gas alone, but $Q_g^{sg}>1$. 
In particular, we have $Q_g^{sg} \xrightarrow[Q_g\rightarrow 0]{} \sqrt{\pi/2}$.
This raises the question of whether the self-gravitating Toomre parameter has a physical meaning or not.
Specifically, does vertical self-gravity acts as a potential stabilising term?
Currently, this is unclear and cannot be analytically demonstrated. 
Indeed, the proof of Toomre's stability criterion relies on the thin disc approximation, which overlooks the vertical stratification of the disc and thus loses information about modified scale heights or differential rotation.
To determine if $Q_i^{sg}$ plays a role in disc stability, rather than $Q_i$, it would be necessary to demonstrate the Toomre stability criterion using a more sophisticated assumption than $\Delta \Phi = 4 \pi G \Sigma \delta(z)$.


\subsection{A closed form for single fluid discs}

When a single fluid is present it is further possible to relate explicitly the midplane density with $k_1$ and the surface density $\Sigma_g$ (see Appendix \ref{subsec: Closure relation for single fluid}).
With this result, the gas density vertical profile is:
\begin{equation}\label{Eq: stratification single fluid}
\displaystyle \rho_g(\vr, z) = \frac{\Sigma_g}{\sqrt{2 \pi} H_g^{sg}} \exp\left[-\frac{1}{2} \left(\frac{z}{H_g} \right)^2 
\displaystyle               - \sqrt{\frac{\pi}{2}} \frac{1}{Q_g^{sg}} \Tf{}{\frac{z}{H_g^{sg}}} \right]
\end{equation}
with:
\begin{equation}\label{Eq: parameters stratification single fluid}
\left\{
\begin{array}{lll}
H_g^{sg} &=& \sqrt{\frac{2}{\pi}} H_g f(Q_g) \\
Q_g^{sg} &=& \displaystyle \sqrt{\frac{\pi}{2}} \frac{Q_g}{f(Q_g)} \\
\end{array}
\right.
\end{equation}
This scale height definition involves particularly the quantity $\sqrt{2/\pi} f(Q_g)$, which weights non-linearly the effect of SG in the stratification.
Particularly, for light discs the above quantity equals unity, while for massive discs it equals $\sqrt{2/\pi} Q_g$.
In Fig. \ref{fig: Gas profile, dust mass negligible} above, we have compared our approximate solution with the BL model and numerical solution in the case of a single fluid.
Across the entire range of $z$ values and Toomre parameters, our model aligns almost perfectly with the numerical solution.
In contrast, the BL model exhibits limitations at small Toomre parameters.
We emphasise that our solution tends asymptotically to the standard Gaussian corrected by an absolute value term, which reinforce the robustness of our solution (see Sect. \ref{sec: Bertin-Lodato and Spitzer models: a discussion}).
Specifically, for strong SG we get: $\log(\rho_g) \xrightarrow[z\rightarrow \infty]{} -\frac{1}{2} (z/H_g)^2 - \frac{\pi}{2} |z|/(Q_g H_g)$.

We believe that only the self-gravitating scale heights can be probed by astronomical observations, which could permit to access indirectly to disc masses.
We address this point in next section.

\section{Connecting theory to observations}
\label{sec: Connecting theory to observations: indirect mass estimation}


In this Sect. we explore the possibility of inferring disc masses through the stratification, and more precisely, the scale heights exhibited in previous sections.
Given the non-standard character of the profiles derived in this work, it is first necessary to define what a dust-to-gas scale height entails in such a case.

\subsection{A generic definition of dust-to-gas scale height}

\begin{figure}
\centering
\includegraphics[width=\hsize]{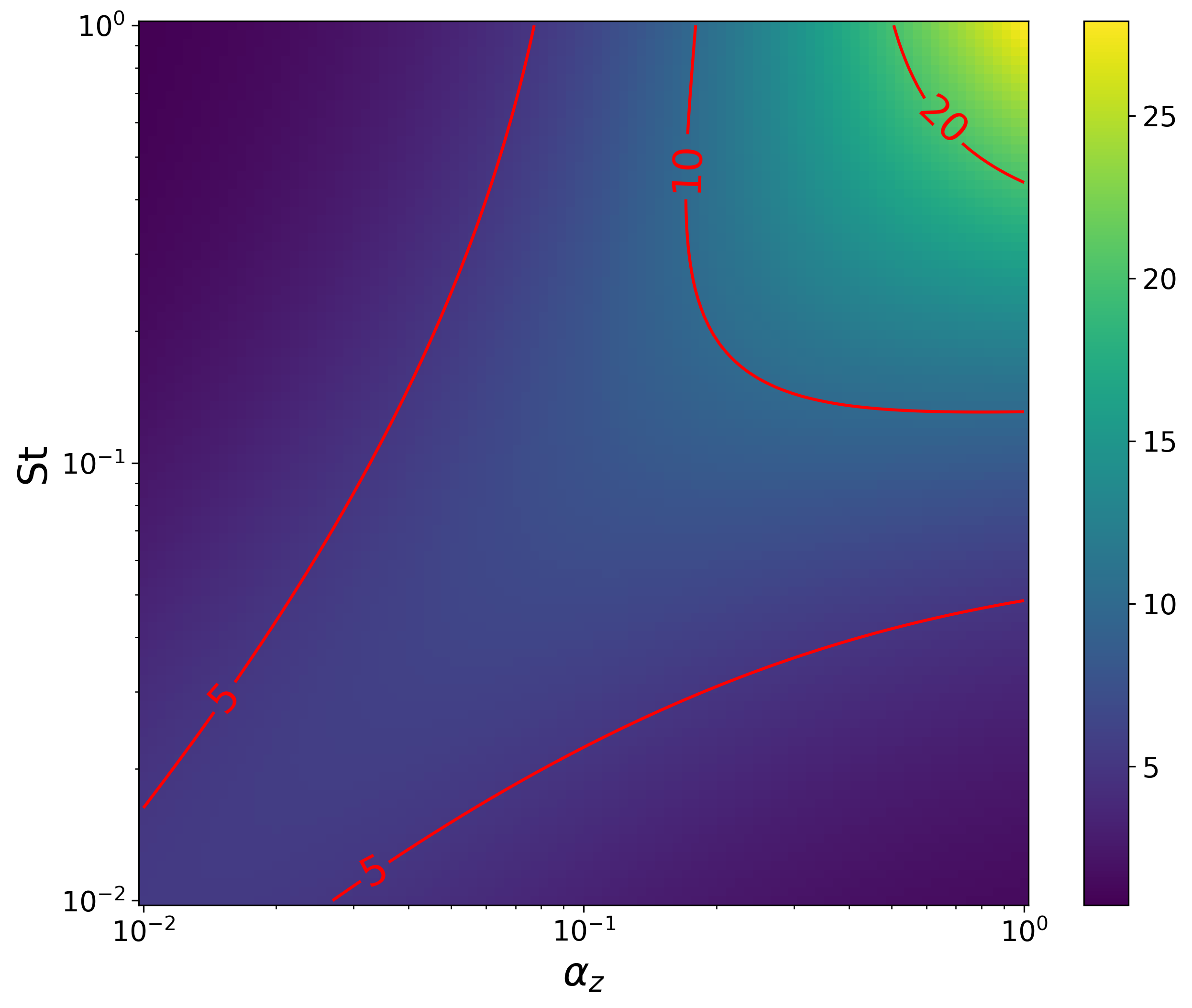}

\caption{Relative difference between the original relation of \citet{1995_dubrulle} and the revised versions that we propose in in Eq. \ref{Eq: revised dubrulle with vertical profile}. 
The red solid lines show contour levels at 5, 10 and 20 \%.
} 
\label{fig: Durbulle relation revised}
\end{figure}

\begin{figure}
\centering
\includegraphics[width=\hsize]{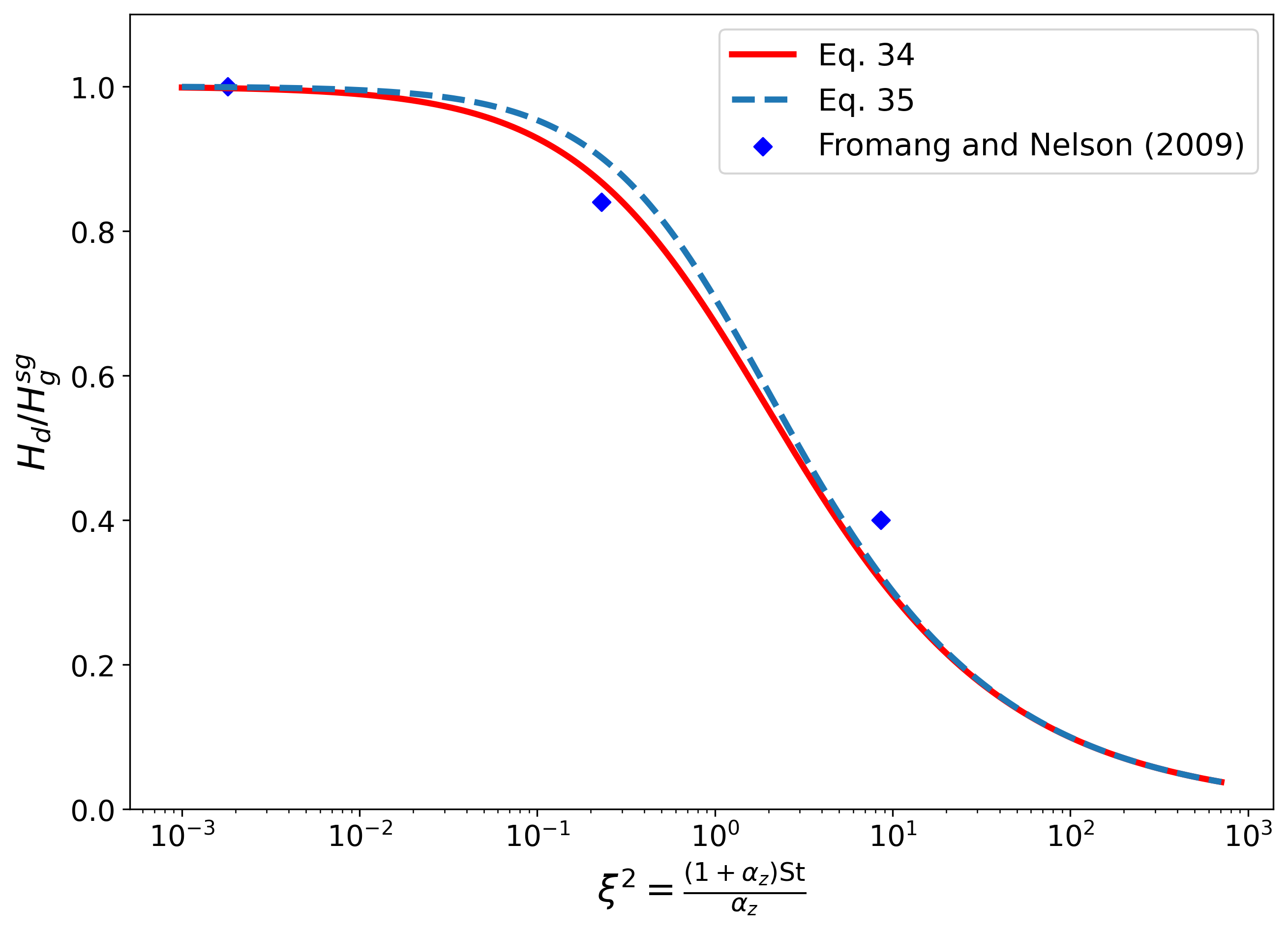}

\caption{Dust-to-gas ratio scale height as a function of the vertical diffusivity and Stokes number when the stopping time has a vertical profile (Eq. \ref{Eq: revised dubrulle with vertical profile}) and when it is constant (Eq. \ref{Eq: revised dubrulle}).
} 
\label{fig: averaged settling}
\end{figure}

The definition of the dust-to-gas scale height ratio is an interesting issue that needs clarification, especially given the complex profiles exhibited in this work.
Finding a general definition of scale heights that applies from Gaussian profiles to non-standard, like the intricate exponential exhibited in Eq. \ref{Eq: No const. stopping time, Bertin, Lodato stratification}, can be challenging. 
Already, the simple $\sech^2$ profile faced this issue, leading \citet[Sect. 2.1]{2021_klahr_schreiber} to revise their initial scale height definition.
In this Sect. we propose a revision of the dust-to-gas scale height that is valid independently of the form of the profile, and therefore applicable even in presence of complex vertical profiles.

First, we investigate a possible definition for the complex profile exhibited in Eq. \ref{Eq: No const. stopping time, Bertin, Lodato stratification}.
Given that both phases are subject to same vertical forces, the dust-to-gas ratio is simply defined as the ratio of the dust diffusivity by the gas sound speed.
However, these last quantities are not simply defined and possibly $z$-dependent for sophisticated profiles. 
For the case of dust, the first step would be to define the correct diffusivity.
To do so it is necessary to write its hydrostatic equilibrium in a symmetric way with respect to the one of gas:
\begin{equation}
\Tcd^2(z) \, \partial_{zz} \ln{\left(\rho_d\right)}  = \displaystyle - \Omega_K^2  - {4 \pi G} \rho_g \,,
\end{equation}
where
\begin{equation}
\displaystyle \Tcd(z)= \frac{c_d(z)}{\sqrt{1+(c_d(z)/\Tcg)^2}} = \sqrt{\frac{X_g(z)}{\xi^2 + X_g(z)}} \Tcg 
\end{equation}
is a $z$-dependent effective dust diffusivity, and the quantity ${X_g(z)=\rho_g(z)/\rho_{g,{\rm mid}}}$ is the normalised gas density.
The next step consists on finding a convenient average method.
A naive approach would consist to integrate above expression from negative to positive infinity and to multiply it with a convenient normalisation factor.
However, this necessarily results in a dust diffusivity that diverges to infinity for $\xi^2=\sto (1+\az)/\az \rightarrow 0$, which is inconsistent. 
To address this, we found more appropriate to use a density weighted average for defining the mean dust diffusivity:
\begin{equation}
\langle \Tilde{c}_i \rangle_{\rho_i} = \int\limits_{-\infty}^{\infty} \rho_i(z) \Tilde{c}_i(z) \, dz \; \Bigg/  \int\limits_{-\infty}^{\infty} \rho_i(z) \, dz
\end{equation}
Finally, we define the general dust-to-gas scale height ratio (valid for any profile of gas and dust) as the ratio of the density weighted dust diffusivity and density weighted gas sound speed:
\begin{equation}\label{Eq: generic definition dust-to-gas scale height}
H_d/H_g^{sg} \coloneqq \langle \Tcd \rangle_{\rho_d} / \langle \Tcg \rangle_{\rho_g} 
\end{equation}
It is important to note that this definition is valid only if both phases are subject to same vertical forces.
A clear benefit of above definition is that it makes possible a generic definition of dust-to-gas scale height ratios valid for complex profiles and at the same time it makes the comparison of dust-to-gas scale height meaningful between different profiles such as Gaussian, hyperbolic secant and intricate exponentials.
At the same time this definition leaves unchanged the standard definition of dust-to-gas scale heights when the diffusion of dust and gas sound speed are space constants.

Now, we apply above definition to the profiles obtained in Sect. \ref{subsec: An approximated solution for any gas disc mass} and Sect. \ref{subsec: the approximated solution}.
According to our prescription, for Eq. \ref{Eq: No const. stopping time, Bertin, Lodato stratification}, the dust-to-gas scale height ratio is:
\begin{equation}\label{Eq: revised dubrulle with vertical profile}
\begin{array}{lll}
H_d/H_g^{sg}
            &=& \displaystyle \int\limits_{-\infty}^{\infty} 
                \displaystyle \sqrt{\frac{G(u)}{\xi^2 + G(u)}} G(u) \exp\left(-\frac{\xi^2}{G(u)}\right) \, du \, \Bigg/  \\
            & & \displaystyle   \displaystyle \int\limits_{-\infty}^{\infty} 
                G(u) \exp\left(-\frac{\xi^2}{G(u)}\right) \, du
\end{array}
\end{equation}
with $G(u)=e^{-{u^2}/2}$.
In Fig. \ref{fig: Durbulle relation revised}, we illustrate the relative difference (in percentage) between the scale height ratio proposed by \citet{1995_dubrulle} and our revised finding in Eq. \ref{Eq: revised dubrulle with vertical profile} for different turbulence strengths and Stokes numbers.
We found notable differences only for strong stirring and large Stokes numbers.
Our investigation revealed that the correction involving the turbulent gas pressure significantly contribute to these deviations.
These conditions might be especially relevant in the context of gravitational instability, as simulations have demonstrated that the $\alpha_S$-parameter can approach values close to unity\citep{2008_kratter}.
This time we apply the definition outlined in Eq. \ref{Eq: generic definition dust-to-gas scale height}, to Eq. \ref{Eq: constant stopping time, SG of gas and dust}.
We get: 
\begin{equation}\label{Eq: revised dubrulle}
\begin{array}{ll}
H_d^{sg}/H_g^{sg}
        &= \langle \Tcd \rangle_{\rho_d} / \langle \Tcg \rangle_{\rho_g} \\
        &= \Tilde{c}_d/\Tilde{c}_g \\
        &= \displaystyle \sqrt{\frac{\az}{\az + \left(1+\az\right) \sto }} \\
        &= \displaystyle \sqrt{\frac{1}{1+\xi^2}}
\end{array}
\end{equation}
In particular, when both phases are subject to the same vertical forces, as is the case here, the above ratio is a physical invariant.
This relation holds true regardless of the presence of SG and is a revision of the formula by \citet{1995_dubrulle}, accounting for how gas turbulence affects not only dust but also its own scale height via a turbulent pressure.
For weak stirring, $\az \ll 1$, we retrieve the result of \citet{1995_dubrulle}.
However, for strong stirring, our revised finding in Eq. \ref{Eq: revised dubrulle} deviates from \citet{1995_dubrulle} relation, similar to how Eq. \ref{Eq: revised dubrulle with vertical profile} deviates from their formula, namely for strong vertical stirring $\az\simeq1$.

Figure \ref{fig: averaged settling} illustrates the dust-to-gas scale height ratio as a function of $\xi^2=(1+\az) \sto / \az$, using both Eq. \ref{Eq: revised dubrulle} and the more generally applicable proposed in Eq. \ref{Eq: revised dubrulle with vertical profile}.
For completeness, we have added the three points that come from \citet{2009b_fromang_nelson} simulations.
Note that while their work originally plotted the dust-to-gas scale height ratio as a function of the Stokes number, we adapted it to be a function of $\xi^2$ based on the rest of their data.
First, we found no significant differences between our two prescriptions.
Therefore, for simplicity, we recommend using the formula provided by Eq. \ref{Eq: revised dubrulle} over that of Eq. \ref{Eq: revised dubrulle with vertical profile}.
Second, we observed that the points from \citet{2009b_fromang_nelson} fit our curves correctly, closely resembling the prescription of \citet{1995_dubrulle}, with deviations appearing only at strong stirrings.
Thus, it is unclear why the authors chose to fit their points with a power law instead of using the known prescription of \citet{1995_dubrulle}.
In summary, we believe that determining the dust-to-gas scale height solely as a function of the Stokes number provides incomplete and biased information.
Instead, it is more appropriate to determine it as a function of $\xi^2=(1+\az) \sto / \az$.

\subsection{Indirect mass estimation}

As shown in Eq.~\ref{Eq: parameters stratification gas and dust}, when accounting for self-gravity the dust and gas scale heights, $H_i^{sg}$, show a symmetric relation with their respective 
$c_i$ terms, and the $k_1$ quantity, which contains information on the disc mass. 
Thus, independent measurements of both the scale height and the temperature, or vertical stirring, of the gas and the dust, respectively, could provide insights into the total disc mass.
In this paragraph, we detail the procedure for the gas alone, which can be easily generalised when dust is present. 
First, it is necessary to obtain $c_g$, which is readily accessible through the gas temperature. 
This allows for the computation of the standard pressure scale height, $H_g$.
The gas temperature can also potentially be determined through dust temperature measurements, assuming thermodynamic equilibrium between the gas and dust. 
Second, the actual vertical scale height, $H_g^{sg}$, which is the real quantity that can be accessed by the data, should be measured by an independent method.
Together, these two scale heights enable the computation of the gas Toomre parameter of the disc:
\begin{equation}\label{Eq: Toomre disc}
Q_g = \sqrt{\frac{\pi}{8}} \frac{2 \eta}{1-\eta^2}
\end{equation}
with $\eta=H_g^{sg}/H_g \leq 1$.
We could link it directly to the local mass of the disc through:
\begin{equation}\label{Eq: Mass disc}
M_{disc}(r) \simeq \sqrt{\frac{8}{\pi}} \frac{1-\eta^2}{\eta} \frac{H_g(r)}{r} M_\odot 
\end{equation}
Employing the above method requires more careful consideration when dealing with dust. 
It is crucial to emphasize that even if the dust temperature is observationally determined, it cannot be directly related to the dust diffusivity: dust is assumed to behave as a pressureless fluid, preventing the association of $\Tcd$ with a temperature. 
Instead, $\Tcd$ is related to the coupling of dust with gas and the level of vertical turbulence. 
As a consequence, only an independent measurement of vertical stirring and knowledge of dust particle sizes would allow the application of the above procedure to dust. 
We will next discuss the feasibility of such observational measurements with current capabilities.

The current capabilities of ALMA, both in terms of sensitivities and angular resolution, have allowed disc temperature measurements to be routinely performed.  
Indeed, in the case where a gas tracer is optically thick, its apparent brightness temperature can be used as a proxy for the local temperature~\citep{Pinte_2023}. 
The brightness temperature of tracers such as $^{12}$CO has been used in combination to the apparent location of its emitting height to retrieve 2D temperature maps~\citep{Dutrey_2017, Pinte_2018, Flores_2021, Leemker_2022}. 
Alternatively, the modelling of the radial emission of observations at several millimeter and centimetre wavelengths also allowed several studies to retrieve the dust midplane temperature in a small sample of discs~\citep[e.g.,][]{Carrasco-Gonzalez_2019, Guidi_2022}.

On the other hand, estimating the dust or gas scale height is observationally more challenging, because these quantities are not directly observable. 
The most direct observable quantity for molecular line is the height of their emission surfaces.
Those can be obtained from geometrical, model-independent methods for observations with sufficient angular resolution to resolve the disc emitting surfaces in the channel maps. This has been employed in about 20 discs with CO observations~\citep{Pinte_2018, Law_2021} and about half of that in other molecular tracers~\citep{ Law_2023, Law_2024, Paneque-Carreno_2024}.
However, directly retrieving the gas scale height from these emission heights is challenging because the latter depends on the gas optical depth. Several attempts have been performed, either assuming a gas scale height from the hydrostatic equilibrium with a midplane temperature based on the stellar luminosity~\citep{Law_2022}, or assuming that the emission heights traces a certain column density~\citep{ Paneque-Carreno_2023}. 
These studies typically find $H_g^{sg}/R\sim0.1$ and apparent height between 2 and 5 times larger than the gas scale heights, but further work is necessary to better characterize the gas scale height. 

Yet, dust scale height estimates are becoming more common. In the optical or infrared, this can be done using detailed radiative transfer modelling of the disc appearance, more efficiently in the case of edge-on discs for which the vertical extent is directly visible~\citep{Burrows_1996, Stapelfeldt_1998, Wolff_2021}. 
The dust scale height estimated at scattered light wavelengths can be used as a proxy for the gas scale height, given that small dust is well coupled to the gas (${\sto<<1}$). 
Alternatively, at longer wavelengths (e.g., millimeter probed by ALMA), grains are affected by vertical settling, and their height can be probed using geometrical effects in their gaps~\citep{2016_pinte, 2022_villenave, Pizzati_2023}. 
This has been performed in various discs, typically estimating millimeter dust scale heights around $H_d^{sg}/R\leq0.01$ in the outer discs, although some systems also show vertically thicker regions~\citep{Doi_Kataoka_2021}. 

Thus, while estimating the dust or gas temperature can be done relatively directly, characterizing the dust or gas scale heights requires more detailed analysis and is not always achievable. 
In the future, combining observational estimates of these tracers could allow to estimate the disc mass in the framework described in this work. 
These results would provide independent and complementary estimates to the disc mass from other methods, such as using observed rotational curve of CO isotopologues~\citep[e.g.,][]{Lodato_2023, Martire_2024}, or thermo-chemical models of different molecular lines~\citep[see][for a review]{2023_miotello_ppvii}. 
To conclude this section, we would like to emphasize that our work has highlighted the necessity of distinguishing between the layering and the temperature of the disc; two concepts that are often mistakenly treated as the same because self-gravity is not taken into account.

%
%
%

\section{Discussion}\label{sec: discussion}

In this Sect. we propose a discussion on the assumptions we made in our model, how our results could affect the trigger of instabilities in PPD and, finally, how they permit an improved treatment of SG in thin disc simulations.

\subsection{Rationale, limitation and improvement of our model}

Scattered light observations of discs, conducted using SPHERE \citep{2018_avenhaus}, allowed the tracing of the gas vertical distribution through small grains. 
Specifically, observations of five discs revealed that the scattered light emission height, typically a few gas scale height above the midplane, follows the relation $H/r=0.162 \left( r / 100 \mbox{AU} \right)^{1.219}$. 
Similarly, observations of the emission heights of millimeter CO isotopologues of more than 10 PPDs allowed to retrieve flared profiles followed by a tapered exponential~\citep{Pinte_2018, Law_2021, Paneque-Carreno_2023}.
$^{12}$CO is typically found to trace layers 2 to 5 scale heights above the midplane, with $z/r < 0.3$.
Consequently, we can safely assume that the gas atmosphere of discs remains flat up to a few hundred AU. 
This effect is even more pronounced for dust, which is expected to be colder and thus highly settled to the midplane, as confirmed by edge-on disc observations \citep{2020_villenave, 2022_villenave}.
Therefore, the realism of the flat disc assumption ensures the validity of the slab approximation, and consequently, the validity of our work. 
However, the slab approximation may fail in regions with strong radial and azimuthal gradients, such as the inner or outer disc, as well as near rings and gaps. 
In the specific case of outer edges, this could result in a factor of 2 difference in the vertical gravitational field generated by the disc.
To address this, \citet{2014_trova} proposed correcting the gravitational field with a rectification factor that accounts for edge effects.
For the outer edge, this factor is:
\begin{equation}
f_{\text{edges}} = \frac{1}{2} + \arctan\left( \varpi_{out} \right) - \varpi_{out} \ln\left( \frac{\varpi_{out}}{\sqrt{1+\varpi_{out}^2}} \right)
\end{equation}
where $\varpi_{out}=\frac{|r_{out}-r|}{2 h(r)}$.
This simple $0^{th}$ order correction permits to decrease the error below 10\%.

Another avenue for improvement would be to revise the assumptions made about the temperature profile or equation of state of the gas.
However, this would increase the complexity of the problem, likely making analytical solutions infeasible and requiring exclusively numerical methods instead.
For example, in the case of a polytropic fluid with $p = \kappa \rho^{1+1/n}$, the equations reduce to the Lane-Emden equations, which are already challenging for spherical symmetries. 
Nevertheless, we found that for $n=1$ and $n=1/2$, there are two solutions that can be expressed in terms of trigonometric and elliptic Weierstrass functions, respectively. 
The derivation of these exact formulas is out of the scope of the present paper.

\subsection{The gravitational collapse of the dusty-gaseous layer}

Beyond a stratification that possibly permits to access other properties of the disc, a better understanding of gas and dust settling would also allow to have a better understanding of the conditions enabling for gravitational instabilities in the dust layer \citep{1969_safronov, 1973_goldreich}. 
According to the studies conducted by \citet{2020_klahr_schreiber, 2021_klahr_schreiber}, the Hill density criterion alone is insufficient to determine the gravitational collapse of pebble clouds.
They found that diffusion, which acts as a pressure-like term, must also be taken into account \citep[Appendix B]{2021_klahr_schreiber}.
This resulted in a Jeans like criterion for dust where they found a critical length above which a self-gravitating pebble cloud can overcome turbulent diffusion and the shear of the star, enabling planetesimal formation.
Specifically, this length equals the vertical scale height of a massive dust layer.
As a consequence, an accurate estimation of the vertical scale height of dust in presence of all gravity terms could permit to better constrain spatial and time regions where dust collapse is supposed to occur.

\subsection{Self-gravity for 2D simulations}

The initial motivation of this study was to determine the correct scale heights for gas and dust in the presence of any SG strength, that could be used as an input to compute accurately SG in thin disc simulations.
Recent work by \citet{rendon_restrepo_2023} highlighted that the commonly used Plummer potential prescription removes the Newtonian character of gravity at mid/short range, leading to inconsistent 2D setups.
Correctly estimating SG in thin discs involves a two-step process. 
The first step, often overlooked, involves determining the vertical hydrostatic equilibrium of the system.
This study aims to address this initial step. 
The resulting stratification is then used in the final phase as an input for vertically integrating all forces, including SG in our specific case.


\section{Conclusion}

In this work, we analytically investigated the layering of a PPD, composed of gas and dust, in the presence of SG.
Specifically, we assumed the gas to be in a turbulent state and the dust to be aerodynamically coupled to it. 

First, we examined the generally applicable case of stopping times with a vertical profile and negligible dust mass.
We derived an exact solution valid for massive gas discs and an approximate solution that bridges Keplerian and massive discs.
Next, we relaxed the assumption on the stopping time, considering a simple case where the stopping time is constant and equal to its midplane value. 
This lead us to find a very accurate and general solution (Eqs. \ref{Eq: stratification gas and dust}-\ref{Eq: parameters stratification gas and dust}) valid for any SG strength.
In particular, we found that under our assumptions, the layering of a disc is governed by three parameters: the gas Toomre parameter, the dust Toomre parameter, and the relative effective gas-to-dust temperature.
We confronted our findings to a physical interpretation and found that the Toomre parameter of a bi-fluid system is the harmonic average of its constituents (Eq. \ref{Eq: Toomre bi-fluid}).
Additionally, in the appendix, we present new exact solutions that can be used to benchmark SG solvers when both gas and dust are present.
We then clarified a possible definition of the dust-to-gas scale height, applicable even to non-standard vertical profiles, when both phases are subject to the same vertical forces.
Specifically, we found that the dust-to-gas ratio is a function of $\xi^2=(1+\az) \sto / \az$, and not solely of the Stokes number.
Finally, we proposed a method for indirectly measuring the mass of discs through the stratification of gas and dust (Eqs. \ref{Eq: Toomre disc}-\ref{Eq: Mass disc}).
In the case of gas, to make this method applicable, it would be necessary to more precisely relate the height of the emitting layer to the gas scale height.

The layerings profiles found in this work are crucial for accurately estimating SG in thin disc simulations.
In future work (currently in preparation), we will leverage our results to discard the smoothing length paradigm.
Finally, observations of thin discs, which possibly indicate low levels of vertical turbulence, might not align well with measured accretion rates.
This discrepancy could be addressed by incorporating SG in gas and dust simulations, for which the stratifications found in this work may be important.
Therefore, in a follow-up paper, we will employ 3D shearing box simulations to study the evolution of a gravito-turbulent dust-gas mixture, a scenario likely occurring in the early ages of PPDs.

\begin{acknowledgements}

Funded by the European Union.
This work is supported by ERC grants (Epoch-of-Taurus, 101043302, PI: O. Gressel) and (DiscEvol, 101039651, PI: G. Rosotti).
Views and opinions expressed are however those of the author(s) only and do not necessarily reflect those of the European Union or the European Research Council. 
Neither the European Union nor the granting authority can be held responsible for them.

\end{acknowledgements}

\bibliographystyle{aa}
\bibliography{bibliography}

\begin{appendix}

\section{Mathematical reference guide}\label{App: mathematical reference guide}

In order to facilitate the reading of this manuscript, we collected non-elementary mathematical results that are necessary for demonstrating several outcomes of this manuscript.

\subsection{Gaussian approximation of hyperbolic profiles}
\label{subsubsec: Gaussian approximation of hyperbolic profiles}

In astrophysical contexts, it is common to approximate the hyperbolic profile with a Gaussian:
\begin{equation}
\displaystyle \frac{1}{2a} \sech^2\left(\frac{x}{a} \right) 
\approx 
\frac{1}{\sqrt{2 \pi} a'} \exp\left({-\frac{1}{2}  \left( \frac{x}{a'} \right)^2} \right)
\end{equation}
with $a'=\sqrt{\frac{2}{\pi}} a$.
The benefit of this approach is twofold: first, it ensures that both functions have the same value at $x=0$; second, it ensures that their integrals from negative infinity to positive infinity are equal.
For $|x| \in [0, 1.6] \, a$ the relative error between both functions is inferior to 9\%.
Outside of the above interval, the error is greater since both functions are not asymptotically equivalent at infinity.
Physically, this is of little importance because both functions approach 0 but mathematically this could impact for the search of more accurate approximate solutions.

\subsection{Special functions}

In this section, we summarized the definitions and properties of several functions that are relevant to our study.

\subsubsection{Modified Bessel functions of the second kind}

For $\Re(z)>0$ and $\Re(\nu)>-\frac{1}{2}$, the integral representation of this function is:
\begin{equation}
K_\nu(z) = \frac{\sqrt{\pi} \, z^\nu}{2^{\nu} \Gamma\left( \nu+\frac{1}{2}\right)} \int\limits_{1}^{\infty} e^{-z t} \left(t^2-1 \right)^{\nu-\frac{1}{2}} dt
\end{equation}
where $\Gamma$ is the Euler gamma function.

\subsubsection{Euler beta function}

For $\Re(x)>0$ and $\Re(y)>0$, this function is defined by:
\begin{equation}
\begin{array}{lll}
\betaf(x,y) &=&\int\limits_0^1 t^{x-1} \left( 1-t \right)^{y-1} dt \\ [10pt]
            &=&\displaystyle \frac{\Gamma(x)\Gamma(y)}{\Gamma(x+y)}
\end{array}
\end{equation}

\subsubsection{Error function}

For $z \in \mathbb{C}$, this function is defined by:
\begin{equation}
\erf(z) = \frac{2}{\sqrt{\pi}} \int\limits_0^{z} e^{-t^2} dt
\end{equation}

\subsection{Integrals}

We collect here the detailed derivation of integrals that permitted us to normalise the different profiles that we propose in this investigation.

\subsubsection{Normalisation of the intricate hyperbolic profile}
\label{subsec: normalisation of the intricate hyperbolic profile}

\begin{equation}
\begin{array}{lll}
\displaystyle  \int_{-\infty}^{\infty} \sech^2(u) & \exp\left[-\xi^2 \cosh^2(u) \right] du &   \\
          & = \displaystyle  \int_{-1}^{1} \exp\left( -\frac{\xi^2}{1-x^2} \right) \, dx   \quad \mbox{with} \quad x=\tanh{u}    &         \\ [10pt]
          & = \displaystyle 2 e^{-\xi^2} \int_{0}^{\infty} \frac{e^{-\xi^2 x^2}}{\left(x^2+1\right)^{3/2}} \, dx &
\end{array}
\end{equation}
Following same steps proposed in \href{https://math.stackexchange.com/questions/2145399/int-11e-frac11-x2dx-can-it-be-computed}{Mathematics StackExchange}\footnote{\url{https://math.stackexchange.com/questions/2145399/int-11e-frac11-x2dx-can-it-be-computed}}, we were able to estimate the quantity in the r.h.s:
\begin{equation}
\begin{array}{lll}
I_1(\xi^2) &=& \displaystyle e^{-\xi^2} \int_{0}^{\infty} \frac{e^{-\xi^2 x^2}}{\left(x^2+1\right)^{3/2}} \, dx  \\ [12pt]
           &=& \displaystyle \frac{\xi^2}{2} e^{-\frac{\xi^2}{2}} \left[ K_1\left( \frac{\xi^2}{2} \right) - K_0\left( \frac{\xi^2}{2} \right) \right]
\end{array}
\end{equation}
where $K_\nu$ is a modified Bessel functions of second kind and of order $\nu$.
Finally, we get:
\begin{equation}
\displaystyle  \int_{-\infty}^{\infty} \sech^2(u) \exp\left[-\xi^2 \cosh^2(u) \right] du = 2 I_1(\xi^2)
\end{equation}

\subsubsection{Normalisation of the hyperbolic profile}
\label{subsec: normalisation of the hyperbolic profile}

\begin{equation}
\begin{array}{lll}
\int\limits_{-\infty}^{\infty} \sech^{2\Txi^2}\left(x\right) \, dx 
         & = &  \displaystyle 
               2 
               \int\limits_{0}^{\infty} 
               \sech^{2 \Txi^2}\left(x\right) \, dx  \\
         & = & \displaystyle 
               2 
               \left[ -\frac{1}{2} \text{B}_{\sech^2(x)} \left(\Txi^2, \frac{1}{2}\right) \right]_{0}^{\infty} \\ [12pt]
         & = & \displaystyle  \text{B}\left(\Txi^2, \frac{1}{2}\right)  
\end{array}
\end{equation}
Here, $\text{B}_x(a,b)$ and $\text{B}(a,b)$ represent the incomplete and complete beta functions, respectively.

\subsection{Taylor series}
\label{subsec: taylor series}



\begin{equation}\label{Eq: beta function asymptotic}
\betaf(x^2, 1/2) \sim \Gamma(1/2) (x^2)^{-1/2} = \frac{\sqrt{\pi}}{x} 
\end{equation}

\section{Stopping time with vertical profile}\label{app: non constant stopping time}

In this Appendix, we detail the derivation of dust vertical layering when star gravity can be neglected, or not, and when the stopping time has a vertical profile.

\subsection{Massive gas disc}\label{app: non constant stopping time, only gas disc contribution}

From the derived gas profile, the dust profile is simply obtained using the transformation defined by Eq. \ref{Eq: reduced eq. SG}.
This results in:
\begin{equation}
\rho_d = A \frac{\Sigma_g}{2 Q_g H_g} \sech^2\left( \frac{z}{Q_g H_g} \right) \exp\left( - \xi^2 \cosh^2\left(  \frac{z}{Q_g H_g} \right) \right)
\end{equation}
where $A$ is an integration constant that we will determine.
This constant is explicitly derived using the definition of the dust column density (Eq. \ref{Eq: column density}):
\begin{equation}
\begin{array}{lll}
\Sigma_d  & = & \displaystyle \int_{-\infty}^{\infty} \rho_d(\vr, z') \, dz' \\ [10 pt]
          & = & \displaystyle \frac{A \Sigma_g}{2} \int_{-\infty}^{\infty} \sech^2(u) \exp\left[-\xi^2 \cosh^2(u) \right] du  \\ [10 pt]
          & = & \displaystyle A \Sigma_g e^{-\xi^2} I_1(\xi^2)
\end{array}
\end{equation}
where we used the results from App. \ref{subsec: normalisation of the intricate hyperbolic profile}.
Taking into account this last result, we finally obtain the profile of dust:
\begin{equation}
\rho_d(\vr,z) = \frac{\Sigma_d(\vr,z)}{ 2 I_1(\xi^2) Q_g H_g} \sech^2\left( \frac{z}{Q_g H_g} \right) \exp\left( - \xi^2 \cosh^2\left(  \frac{z}{Q_g H_g} \right) \right)
\end{equation}

\subsection{Gas disc and star contribution}\label{app: non constant stopping time, star + gas disc contribution}

When the gas vertical profile obeys the BL layering, the dust profile is:
\begin{equation}
\begin{array}{cc}
\rho_d(\vr,z) & = \displaystyle A \frac{\Sigma_g}{\sqrt{2\pi} H_g^{sg}} \exp\left(-\frac{1}{2} \left({z}/{H_g^{sg}}\right)^2\right) \\ [12pt]
              &   \qquad  \exp\left[ - \xi^2 \exp\left(\frac{1}{2} \left(\frac{z}{H_g^{sg}}\right)^2\right) \right]
\end{array}
\end{equation}
Similarly to Appendix \ref{app: non constant stopping time, only gas disc contribution}, above density must be integrated for getting the integration constant $A$, which leads to:
\begin{equation}
\Sigma_d = \displaystyle A \frac{\Sigma_g}{\sqrt{2\pi}} \int\limits_{-\infty}^{\infty} e^{-\frac{u^2}{2}} \exp\left(-\xi^2 \exp\left(\frac{u^2}{2}\right)\right) \, du
\end{equation}
We were not able to find a closed-form for the integral in the r.h.s of above equation in terms of standard mathematical functions. 
Nevertheless, using the approximation $\exp\left( -{x^2}/{2} \right) \approx \sech^2\left(\sqrt{2/\pi} x \right)$, we were able to express it with the function $I_1$ defined in Appendix \ref{subsec: normalisation of the intricate hyperbolic profile} and used in Appendix \ref{app: non constant stopping time, only gas disc contribution}, as:
\begin{equation}
\begin{array}{cc}
\rho_d(\vr,z) & = \displaystyle \frac{\Sigma_d}{\sqrt{2\pi} I_1(\xi^2) H_g^{sg}} \exp\left(-\frac{1}{2} \left({z}/{H_g^{sg}}\right)^2\right) \\ [12pt]
              &   \qquad  \exp\left[ - \xi^2 \exp\left(\frac{1}{2} \left(\frac{z}{H_g^{sg}}\right)^2\right) \right]
\end{array}
\end{equation}

\section{The Bertin-Lodato approach}\label{app: Bertin-Lodato approximation}

Here we aim to propose a derivation of the biased Gaussian approximation proposed by \citet{1999_Bertin_Lodato} to Eq. \ref{Eq: constant stopping time, SG of gas} with our notations.
The power and simplicity of this method is that all SG information in incorporated in a modified scale height.
We start by the demonstration of the BL profile.

\subsection{Demonstration}\label{subsec: Bertin-Lodato model demonstration}

When the disc is only made of gas, \citet{1999_Bertin_Lodato} demonstrated the possibility of obtaining an approximate solution that ensures a smooth transition between Keplerian and massive discs. 
Their approximate solution, following a Gaussian profile, relies on the premise that for weak SG, the density's vertical profile is precisely Gaussian.
Conversely, for strong SG, the vertical layering adopts a $\sech^2(x)$ profile, closely resembling a Gaussian through a mere rescaling of the variable x.
As a consequence, the gas density obey next Gaussian distribution:
\begin{equation}\label{Eq: Demonstration gas gaussian}
\rho_g(\vr,z)  = \displaystyle \frac{\Sigma_g}{\sqrt{2\pi} H_{g}^{sg}} \exp\left[-\frac{1}{2} \left(z/H_{g}^{sg}\right)^2\right] 
\end{equation}
where $H_{g}^{sg}$ is a modified gas scale height.
Notably, this scale heights is expected to encompass information related to the disc SG, prompting to make following coherent assumption:
\begin{equation}\label{Eq: scaling H_g}
\displaystyle H_g^{sg} = \sqrt{\frac{2}{\pi}} H_g f(Q_g) 
\end{equation}
where the function $f$ is an unknown that need to be found self-consistently later in our reasoning.
Particularly, it is expected that $Q_g$ will permit a connection between the Keplerian disc and the Spitzer disc.

We substitute Eq. \ref{Eq: Demonstration gas gaussian} into Eq. \ref{Eq: constant stopping time, SG of gas}. 
At the $0^{th}$ order in $z$, we obtain:
\begin{equation}\label{Eq: Taylor expansion 0 order}
\frac{\pi}{2} \frac{1}{H_g^2 f^2(Q_g)} = 
 \frac{1}{H_g^2}
+
\frac{2}{f(Q_g)} \frac{1}{H_g^2} 
\frac{1}{Q_g}
\end{equation}
After reorganising all terms in this relation, we get:
\begin{equation}\label{Eq: second order Eq f}
f(Q_g)^2 + \frac{2 f(Q_g)}{Q_g}  - \frac{\pi}{2} = 0
\end{equation}
We highlight that $f$ is a function solely dependent on the variable $Q_g$, as initially hypothesised, confirming the robust self-consistency of the initial assumption.  
The second-order Eq. \ref{Eq: second order Eq f} has the unique acceptable solution:
\begin{equation}
f(Q_g) = \frac{1}{Q_g} \left( \sqrt{1+ \frac{\pi}{2} Q_g^2} - 1  \right)
\end{equation}
This solution respects the limit for weak SG, 
\begin{equation}
\lim\limits_{Q_g \rightarrow \infty} H_g^{sg} =  H_g
\end{equation}
but fails to meet the constraint for strong SG, 
\begin{equation}
\lim\limits_{Q_g \rightarrow 0} H_g^{sg}  = Q_g H_g
\end{equation}
This discrepancy was easily rectified by \citet{1999_Bertin_Lodato} by adjusting $Q_g$ in the function $f$ by a constant factor: $Q_g \rightarrow 4 / \pi Q_g$. 
We found that this correction maintains the constraint at the weak SG limit unchanged.
Finally, the expression of $f$ is:
\begin{equation}
f(Q_g) = \frac{\pi}{4 Q_g} \left( \sqrt{1+ \frac{8}{\pi} Q_g^2} - 1  \right)
\end{equation}
This expression of $f$ ensures the validity of the gas scale height across weak to strong SG.
In next Sect. we propose to address the issue related to the $\pi/4$ factor.

\subsection{Consistency of the Gaussian approximation}\label{subsec: consistency of the gaussian biased approximation}

The hydrostatic equilibrium of a massive self-gravitating gas disc is given by:
\begin{equation}\label{Eq: SG of gas}
\displaystyle c_g^2 \partial_{zz} \ln{\left(\rho_g\right)}  = - {4\pi G} \rho_g
\end{equation}
The well-known solution to this equation, as described by \citet{1942_spitzer}, is:
\begin{equation}
\rho_g(r,z) = \displaystyle \frac{\Sigma_g}{2 Q H}  \sech^{2} \left( \frac{z}{Q H} \right)
\end{equation}
For simplicity, it is common to approximate this hyperbolic profile with a Gaussian profile (see Appendix \ref{subsubsec: Gaussian approximation of hyperbolic profiles}):
\begin{equation*}
\rho_g(r,z) = \displaystyle \frac{\Sigma_g}{2 Q H}  \exp\left( - \frac{\pi}{4} \left(\frac{z}{Q H} \right)^2 \right)
\end{equation*}
However, when this Gaussian approximation is substituted into Eq. \ref{Eq: SG of gas}, a discrepancy of a factor $\pi/4$ appears between the left and right-hand side terms at 0$^{th}$ order in $z$.
This discrepancy arises due to non-linear effects that are not captured by the Gaussian approximation.
In simple words, the Gaussian biased profile is an accurate approximation of the analytic solution (that requires it to be known) but it cannot be used directly in the primitive equation \ref{Eq: SG of gas}.
Therefore, to address this issue it is necessary to rectify simply all SG terms by aforementioned factor $\pi/4$.


\subsection{Discussion}

The method exhibited in this appendix is however not generally applicable for bi-fluids when their temperatures start to differ because of the non-linear nature of the equations.
Indeed, we applied this method to Eq. \ref{Eq: unique Liouville eq} and found modified scale heights of gas and dust that depend on a general Toomre's parameter:
\begin{equation}
\TQ=\left( \frac{1}{Q_g} + \frac{1}{Q_d}\right)^{-1}
\end{equation}
Even though this result seems coherent with the underlying physics, we think that it is incorrect because any definition of the Toomre's parameter of a bi-fluid should also be dependant on the relative temperature, $\Txi$, which is not the case here.
Therefore, it is necessary to find a more general and alternative method valid for any $\Txi$, that we present in Appendix \ref{app: A general approximated solution}.

\section{Particular solutions when $\tau=\tau_{{\rm mid}}$}\label{app: constant stopping time}

In this section, we compile and organise known, and new, analytic solutions to Eq. \ref{Eq: constant stopping time, SG of gas and dust} for relevant specific cases.
These particular exact solutions will enable us to validate the legitimacy of the general approximate solution proposed in Sect. \ref{app: A general approximated solution}.

\subsection{Keplerian disc, $Q_g, Q_d \gg 1 $}\label{app: constant stopping time, keplerian}

When the SG contribution from gas and dust can be neglected, the well known Gaussian stratification is obtained \citep[Eqs. 15 and 238]{2019_armitage}:
\begin{equation}\label{Eq: stratification gas and dust keplerian}
\left\{
\begin{array}{cc}
\rho_g(\vr,z) & = \displaystyle \frac{\Sigma_g}{\sqrt{2\pi} H_{g}} \exp\left[-\frac{1}{2} \left(z/H_{g}\right)^2\right] \\ [8pt]
\rho_d(\vr,z) & = \displaystyle \frac{\Sigma_d}{\sqrt{2\pi} H_{d}} \exp\left[-\frac{1}{2} \left(z/H_{d}\right)^2\right] 
\end{array}
\right.
\end{equation}
where $H_i = \Tilde{c}_i/\Omega_K$ is the standard definition of the scale height of phase $i$.

\subsection{Massive gas disc, $Q_g \ll 1$ and $Q_d \gg Q_g$}\label{app: constant stopping time, massive gas disc}

When the SG of gas dominates over the contributions from the star and dust, we obtain following vertical profiles:
\begin{equation}\label{Eq: exact high gas mass, constant stopping time incomplete}
\left\{
\begin{array}{lll}
\rho_g(\vr,z)&=&\displaystyle \frac{\Sigma_g}{2 Q_g H_g}  \sech^{2}\left(\frac{z}{Q_g H_g}\right) \\ [10pt]
\rho_d(\vr,z)&=&\displaystyle 
A 
\left(\frac{\Sigma_g}{2 Q_g H_g} \right)^{\Txi^2}
\sech^{2 \Txi^2}\left( \frac{z}{Q_g H_g} \right) \\
\end{array}
\right.
\end{equation}
where $A$ is an integration constant.
Notably, for the gas, we obtained the well known Spitzer profile \citep{1942_spitzer}.
The integration constant for dust is determined using the definition of dust column density (Eq. \ref{Eq: column density}):
\begin{equation}
\begin{array}{lll}
\Sigma_d & = & \displaystyle A
               \left(\frac{\Sigma_g}{2 Q_g H_g} \right)^{\Txi^2}
               Q_g H_g
               \int\limits_{-\infty}^{\infty} \sech^{2\Txi^2}\left(x\right) \, dx                  \\
         & = & \displaystyle A
               \left(\frac{\Sigma_g}{2 Q_g H_g} \right)^{\Txi^2}
               Q_g H_g \text{B}\left(\Txi^2, \frac{1}{2}\right)                                              \\
\end{array}
\end{equation}
where we used the integral from Appendix \ref{subsec: normalisation of the hyperbolic profile}.
Here, $\text{B}(a,b)$ represent the complete beta function.
This final result allows us to obtain the complete vertical profile of both phases:
\begin{equation}\label{Eq: exact high gas mass, constant stopping time full}
\left\{
\begin{array}{lll}
\rho_g(\vr,z)&=&\displaystyle \frac{\Sigma_g}{2 Q_g H_g}  \sech^{2}\left(\frac{z}{Q_g H_g}\right) \\ [10pt]
\rho_d(\vr,z)&=&\displaystyle \frac{\Sigma_d}{2 Q_g H_g} \frac{2}{\text{B}(\Txi^2, 1/2)} \sech^{2 \Txi^2}\left( \frac{z}{Q_g H_g} \right) \\
\end{array}
\right.
\end{equation}
To our knowledge the stratification of dust is new.

A physical insight of these relations is found in the limit of high gas temperatures, $\Txi^2 \rightarrow \infty$, which allowed us to find the scale heights of both fluids under the form:
\begin{equation}
H_i^{sg} \approx Q_g H_i
\end{equation}
where we used Eq. \ref{Eq: beta function asymptotic} for the definition of dust scale height.
The presence of the gas Toomre's parameter in both scale heights relations simply indicates that the layering of gas, and respectively dust, is determined by the balance between gas pressure, and respectively dust vertical stirring, and the self-gravity generated by the gas disc alone. 
We anticipate this solution to be correct in the evolution stages and regions of PPDs where the disc mass is dominated by the gas phase, which is the case for Class 0/I discs and the outer regions of PPDs.

\subsection{Massive dust disc, $Q_d \ll 1$ and $Q_g \gg Q_d$}\label{app: constant stopping time, massive dust disc}

The vertical layering of a PPD whose gravitational contribution is dominated by dust mass is simply obtained using same method as in Sect. \ref{app: constant stopping time, massive gas disc}:
\begin{equation}\label{Eq: massive dust disc exact}
\left\{
\begin{array}{ccl}
\rho_g(r,z) & = &\displaystyle \frac{\Sigma_g}{2 Q_d H_d} \frac{2}{\text{B}(1/\Txi^2,1/2)} \sech^{2/\Txi^2}\left( \frac{z}{Q_d H_d} \right) \\ [10pt]
\rho_d(r,z) & = &\displaystyle \frac{\Sigma_d}{2 Q_d H_d} \sech^{2}\left(\frac{z}{Q_d H_d}\right)
\end{array}
\right.
\end{equation}
The relative temperature is $\Txi=\Tcg/\Tcd$.
This time we retrieve the dust vertical profile proposed by \citet{2020_klahr_schreiber, 2021_klahr_schreiber} while the gas density profile is modified.
This profile is also new.

Symmetrically to previous case, the SG of the solid material disc is compensated by the gas pressure and dust vertical diffusion.
This statement can be expressed mathematically for each scale height under the form: 
\begin{equation}
H_i^{sg} \approx Q_d H_i
\end{equation}
We think that the stratification described by Eq. \ref{Eq: massive dust disc exact} should be used in regions highly depleted in gas.
This could correspond to debris discs or, more generally, in regions where photoevaporation already occurred.

\subsection{Massive disc and perfect coupling, $\Txi=1$}\label{app: constant stopping time, same velocity dispersion}

Here, we examine the scenario where gas and dust exhibit the same layering, which corresponds to the regime of well-coupled grains, $\sto\ll1$.
Although this case is simpler compared to the previous ones, we have chosen to present this solution because, from an analytical perspective, it allows for an additional stratification.
This is valuable for constructing a consistent approximated solution later on.
The solution profile for this specific case is:
\begin{equation}
\left\{
\begin{array}{ccl}
\rho_g(r,z) & = &\displaystyle \frac{\Sigma_g}{2 Q H}  \sech^{2} \left( \frac{z}{Q H} \right) \\ [10pt]
\rho_d(r,z) & = &\displaystyle \frac{\Sigma_d}{2 Q H}  \sech^{2} \left( \frac{z}{Q H} \right)    
\end{array}
\right.
\end{equation}
where $H=\tilde{c}_i/\Omega_K$ and $Q = \left( \frac{1}{Q_g} + \frac{1}{Q_d}  \right)^{-1}$ is the Toomre's parameter of both species.
Remarkably, the scale height is uniquely defined by the cumulated mass of both species.
We can rewrite this result in next manner:
\begin{equation}
H_i^{sg} \approx \left( \frac{1}{Q_g} + \frac{1}{Q_d}  \right)^{-1} H_i
\end{equation}
Notably, when the mass of the gas disc, or dust disc, prevails, we retrieve the scale heights of Sects. \ref{app: constant stopping time, massive gas disc} and \ref{app: constant stopping time, massive dust disc}, respectively, when $\Txi=1$.

\subsection{Summary}

In this section, we have presented and organised specific solutions to equation Eq. \ref{Eq: constant stopping time, SG of gas and dust}.
In particular, we have discussed the standard Gaussian solution, which is valid when the stellar gravity dominates (Sect. \ref{app: constant stopping time, keplerian}), two new solutions applicable when the mass of gas or dust prevails (Sects. \ref{app: constant stopping time, massive gas disc} and \ref{app: constant stopping time, massive dust disc}), and a final solution valid when both phases share same relative temperature (Sect. \ref{app: constant stopping time, same velocity dispersion}).

Through inductive reasoning, this classification has enabled us to predict that the scale height of gas and dust, in the presence of stellar gravity and when both phases SG is incorporated, should take the form:
\begin{equation}\label{Eq: prediction scale heights}
\begin{array}{ll}
H_g^{sg}  & \approx  h(Q_g, Q_d, \Txi) H_g  \\ [4pt]
H_d^{sg}  & \approx  h(Q_g, Q_d, \Txi) H_d
\end{array}
\end{equation}
where the function $h$ satisfies:
\begin{equation}\label{Eq: function h properties}
h(Q_g, Q_d, \Txi) = 
\left\{
\begin{array}{ll}
1   & \mbox{if star gravity prevails} \\
Q_g & \mbox{if gas mass prevails} \\
Q_d & \mbox{if dust mass prevails} \\
 \left( \frac{1}{Q_g} + \frac{1}{Q_d}  \right)^{-1} 
    & \mbox{if same relative temperature}
\end{array}
\right.
\end{equation}
This expression for $h$ permits to satisfy the limit cases investigated in four previous subsections.
Any general solution, or approximated solution, must respect the structure of the scale heights provided by Eq. \ref{Eq: prediction scale heights}.
In the next section, we propose an approximate solution that respect these constraints.

\section{A general approximated solution}
\label{app: A general approximated solution}

In this appendix we are concerned by an accurate and general resolution of Eq. \ref{Eq: unique Liouville eq}.
As for the Bertin-Lodato approximation we will make use of a Gaussian like approximation, even though more sophisticated.
For simplicity we used next variable substitutions:  $\Tz=z/H_g$, $a=\frac{\Omega_K^2}{ 4\pi G \rho_{g,{\rm mid}}}$, $b=\frac{\Omega_K^2}{4 \pi G \rho_{d,{\rm mid}}}$ and $X_g=\rho_g/\rho_{g,{\rm mid}}$.
This permits, to rewrite above equation in a simpler manner:
\begin{equation}\label{Eq: Liouville eq. rectified and normalised}
\partial_{\Tz \Tz} \ln{\left(X_g\right)} = -1 - \frac{1}{a} X_g - \frac{1}{b} X_g^{\Txi^2} \\
\end{equation}
Accounting for the discussion of Sect. \ref{sec: Bertin-Lodato and Spitzer models: a discussion}, we know that the solution should adopt a Gaussian like profile at infinity.
Motivated by the structure of Eq. \ref{Eq: Liouville eq. rectified and normalised} and last comment, we propose to adopt an iterative method for searching for a solution under the form:
\begin{equation}
X_g=\exp\left[ -\frac{1}{2} \Tz^2 - \varphi_i(\Tz) \right]
\end{equation}
where $\varphi_i$ is a function that would be refined at each iteration.
Accounting that close to the midplane $X_g=1$, we get for the first iteration:
\begin{equation}
-1 - \varphi_1'' = -1 -\frac{1}{a} - \frac{1}{b} 
\end{equation}
whose solution is: 
\begin{equation}
\varphi_1(\Tz) = \frac{1}{2} \left( \frac{1}{a} + \frac{1}{b}  \right) \Tz^2
\end{equation}
For the second iteration we get:
\begin{equation}
\begin{array}{lll}
\varphi_2'' = \frac{1}{a} \exp(-\frac{1}{2} \left(\frac{\Tz}{k_1} \right)^2 ) 
+ \frac{1}{b} \exp(-\frac{1}{2} \left(\frac{\Tz}{k_1/\Txi} \right)^2 ) 
\end{array}
\end{equation}
where $k_1 =  1/\sqrt{1+\frac{1}{a} + \frac{1}{b}}$.
The solution of the second iteration is:
\begin{equation}
\varphi_2(\Tz) =\displaystyle  \frac{1}{a} \Tf{k_1}{\Tz} + \frac{1}{b} \Tf{k_1/\Txi}{\Tz} \\
\end{equation}
with:
\begin{equation}\label{Eq: Theta definition}
\Tf{k}{\Tz}    = \displaystyle k^2 \left[ \exp\left[-\frac{1}{2} \left(\frac{\Tz}{k}\right)^2 \right] 
                       + \sqrt{\frac{\pi}{2}} \frac{\Tz}{k} \text{erf}\left(\frac{\Tz}{\sqrt{2}k}\right) - 1 \right]
\end{equation}
We couldn't go further in the iterations because it was not possible for us to integrate $\exp(-\frac{1}{2} \Tz^2 - \varphi_2(\Tz))$.
Nevertheless, this approximate solution is very accurate for almost all regimes of SG of gas and dust.
Accounting for the discussion in Appendix \ref{subsec: consistency of the gaussian biased approximation}, all SG terms should be rescaled by a factor $\pi/4$.
Therefore, we do the substitution: $\frac{1}{a} \rightarrow \frac{\pi}{4a}$ and $\frac{1}{b} \rightarrow \frac{\pi}{4b}$.
We indeed, checked that this simple transformation highly improves the accuracy of our approximated solution.

In summary, with the initial notations the approximated solution that we propose reads:
\begin{equation}\label{Eq: sol intermediate}
\left\{
\begin{array}{ll}
\displaystyle 
\rho_g(z) = \rho_{g,{\rm mid}} \exp\left[
-\frac{1}{2} \left(\frac{z}{H_g} \right)^2 \right.
            & \displaystyle - \frac{\pi^2 G \rho_{g,{\rm mid}}}{\Omega_K^2} \Tf{k_1}{\frac{z}{H_g}}                 \\
            & \displaystyle \left. - \frac{\pi^2 G \rho_{d,{\rm mid}}}{\Omega_K^2} \Tf{k_2}{\frac{z}{H_g}}  \right] \\
\rho_d(z) = \rho_{d,{\rm mid}} \left[ \rho_g/\rho_{g,{\rm mid}} \right]^{\Txi^2}
\end{array}
\right.
\end{equation}
with the function $\Theta_k$ defined in Eq. \ref{Eq: Theta definition} and:
\begin{equation}\label{Eq: k1 and k2 definition}
k_1 = 1\Big/ \sqrt{1+\frac{\pi^2 G}{\Omega_K^2}\left(\rho_{g,{\rm mid}} + \rho_{d,{\rm mid}} \right)} \quad \text{and} \quad k_2=k_1/\Txi
\end{equation}
However, this solution does not involve measurable quantities like surface densities.
It also seems obvious that a closure condition is missing, since midplane densities should explicitly be dependent on $k_1$, which also depends on mid-plane densities.
As a consequence, our next aim is to express, when possible, the midplane densities as a function of surface densities and other physical parameters.
We start by treating the simpler single fluid case.

\subsection{Closure relation for single fluid}
\label{subsec: Closure relation for single fluid}

To compute midplane values, one could employ the method used several times in previous sections, namely integrating the gas and dust density from negative infinity to positive infinity to obtain the associated surface densities. 
However, for the complex relation of Eq. \ref{Eq: sol intermediate}, this approach requires approximations that result in non-linear equations involving Gaussian and error functions, which are not easy to solve. 
Instead, we adopted a simpler and more ingenious method that involves substituting the solution we found for the gas (Eq. \ref{Eq: sol intermediate}) into the original equation we initially needed to solve:
\begin{equation}
\Tcg^2 \partial_z \ln{\rho_g} = -\Omega_K^2 z - 2 \pi G  \int\limits_{-z}^{z} \rho_g(u) \, du 
\end{equation}
When substituting Eq. \ref{Eq: sol intermediate}, and taking the limit as $z\rightarrow \infty$, we obtain:
\begin{equation}
\rho_{g, {\rm mid}}  = \frac{\Sigma_g }{\sqrt{2 \pi} H_g k_1} 
\end{equation}
This is a self-consistent condition on the midplane density. 
Indeed, $k_1$ is itself dependent on this quantity (see Eq. \ref{Eq: k1 and k2 definition}), leading to a second-order equation similar to Eq. \ref{Eq: second order Eq f}, whose acceptable solution is:
\begin{equation}
k_1 = \sqrt{\frac{2}{\pi}} f(Q_g)
\end{equation}
where we already applied the $\pi/4$ factor correction to SG terms.

\subsection{Closure relation for bi-fluids}

Adopting the procedure of previous Sect. to a mixture of gas and dust, we obtained:
\begin{equation}
\rho_{g, {\rm mid}} + \frac{\rho_{d,{\rm mid}}}{\Txi} = 
\frac{1}{\sqrt{2 \pi} H_g k_1} \left( \Sigma_g + \Sigma_d \right)
\end{equation}
We couldn't find a second constraint relating these different quantities and permitting to obtain $k_1$ explicitly.
Indeed, simple approximations showed that the volume densities ratios of gas and dust are a complex function of $\Sigma_g$, $\Sigma_d$ and $\Txi$, which does not help much.

\end{appendix}

\end{document}